\documentclass[a4paper,UKenglish,letter]{lipicsown}
\usepackage{url}
\usepackage{color}
\usepackage[usenames,dvipsnames,svgnames,table]{xcolor}
\hypersetup{colorlinks   = true,
     citecolor    = black,
     urlcolor     = violet,
     linkcolor    = blue}
\usepackage{amsfonts}
\usepackage{amssymb}
\usepackage{float}
\usepackage{enumerate}
\usepackage{tikz}
\newtheorem{exmp}[theorem]{Example}
\usepackage{amsmath}
\usepackage{mathtools}

\newlength\myindent 
\usepackage{enumerate}

\title{Addition Machines, Automatic Functions and Open Problems of Floyd and
  Knuth}


\author[1]{Sanjay Jain$^1$, Xiaodong Jia$^1$, Ammar Fathin Sabili}
\author[2]{Frank Stephan}
\affil[1]{School of Computing, National University of Singapore,
13 Computing Drive, Block COM1, Singapore 117417,
Republic of Singapore, \texttt{sanjay@comp.nus.edu.sg}; 
\texttt{ammar@comp.nus.edu.sg}; \texttt{j.xiaodong@u.nus.edu};
S.~Jain was supported in part by the Singapore Ministry of Education
Academic Research Fund Tier 2 grant MOE2019-T2-2-121 / R146-000-304-112
as well as NUS Provost Chair grant C252-000-087-001.}
\affil[2]{Department of Mathematics and School of
Computing, National University of Singapore,
10 Lower Kent Ridge Road, Block S17, Singapore 119076,
Republic of Singapore, \texttt{fstephan@comp.nus.edu.sg};
F.~Stephan was supported in part by the Singapore Ministry of Education Academic
Research Fund Tier 2 grant MOE2019-T2-2-121 / R146-000-304-112.}

\authorrunning{S.~Jain, X.~Jia, A.F.~Sabili and F.~Stephan}

\titlerunning{Addition Machines, Automatic Functions and Open Problems
  of Floyd and Knuth}

\date{\today}


\begin{document}
\maketitle

\begin{abstract}
\noindent
Floyd and Knuth investigated in 1990 register machines which can
add, subtract and compare integers as primitive operations. They
asked whether their current bound on the number of registers
for multiplying and dividing fast (running in time linear in the size
of the input) can be improved and whether one can output fast the powers
of two summing up to a positive integer in subquadratic time.
Both questions are answered
positively. Furthermore, it is shown that every function computed
by only one register is automatic and that automatic functions with
one input can be computed with four registers in linear time; automatic
functions with a larger number of inputs can be computed with 5 registers
in linear time. There is a nonautomatic function with one input which
can be computed with two registers in linear time.
\end{abstract}

\bigskip
\noindent
{\bf Table of Contents}
\begin{enumerate}
\item Introduction \dotfill \pageref{sec:intro}
\item On the Methods and Notions in this Paper
        \dotfill \pageref{sec:methods}
\item Solving Open Problems (2) and (5) of Floyd and Knuth
        \dotfill \pageref{sec:twofive}
\item Open Problems (3) and (4) and Multivariate Oh Notations
        \dotfill \pageref{sec:threefour}
\item Regular Languages and Automatic Functions
        \dotfill \pageref {sec:autofunc}
\item Sample Programs \dotfill \pageref{sec:sampleprog}
\item Conclusion \dotfill \pageref{sec:conclu}
\item References \dotfill \pageref{sec:thebib}
\end{enumerate}

\newpage

\section{Introduction}  \label{sec:intro}

\noindent
Hartmanis and Simon \cite{HS74} showed that register machines which
can add and multiply and perform bitwise Boolean operations
in unit time and which can hold arbitrarily big numbers can
solve {\bf NP}-hard problems (and beyond) with polynomially many
primitive steps. Though this model is powerful, mathematicians
and computer scientists have nevertheless seen quite some usefulness
in determining for arithmetic circuits using $+$ and $*$ how many
operations and what depth they need to compute natural concepts
like permanents and determinants over integers and other rings.
Early work in this area is surveyed by Pippenger \cite{Pi81}.
A more recent
sample result is by Agrawal and Vinay \cite{AV08} showing that
circuits of depth four (with multiple-fanin for both arithmetic
operations) can compute much more than those of depth 3.
Furthermore, blackbox algorithms for polynomial identity testing
use algebraic operations as their primitives, with some assumptions
on the ring used in which these computations are carried out;
for example one might try to minimise the amount of randomness
used in such an algorithm \cite{AB03}. These studies address various
models like the usage of algebraic operations on real numbers, the
study of operations on the integers and on finite fields.
However, when considering the full functionality of arithmetics
in computer programs or circuits, the model is too powerful when
using infinite rings like the integers.
In contrast to this, Hartmanis and Simon \cite{HS74} showed
that register machines which have only addition, subtraction and comparison
instructions on the integers can carry out arbitrary polynomial time operations
in polynomial time but not more, so they are a realistic model of
computation with primitive steps that are more comprehensive than
those of Turing machines or counter machines. Due to the
unit costs of addition, some operations became faster, for
example multiplication can be done in $O(n)$ instead of the fastest known
$O(n \log n)$, which counts multitape Turing machine steps
and was recently discovered by Harvey and van der Hoeven \cite{HH19}.
By definition, addition and subtraction are $O(1)$ instead of $O(n)$.
On the other hand, Stockmeyer \cite{St76} showed that
even if a machine can add, subtract and multiply
in unit time, it cannot compute the following functions in $o(n)$ steps:
$x \mapsto x/2$, $x \mapsto (x \mod 2)$; the same lower bound
also holds for addition machines.
Indeed, for many regular sets, addition machines need time $\Theta(n)$
similar to Turing machines which must often inspect all digits to decide
the membership.

Floyd and Knuth \cite{FK90} systematically
studied register machines which can add, subtract and compare and called
them ``addition machines''; in the following text, addition machines and
register machines are used synonymously. Knuth recalled this work
in Floyd's obituary \cite{Kn04} as one of the joint works he enjoyed a lot
in the later stage of their collaboration. They found that addition machines
form a natural model and they provided various algorithms for arithmetic
on them, in particular as Floyd and Knuth looked for alternatives to the
usual Turing machine models with their tiny primitive steps.
Anisimov \cite{An95,An99} studied the idea of Floyd and Knuth
of using ideas borrowed from Fibonacci numbers for implementing arithmetic
on large numbers with addition and investigated it thoroughly; later,
he and Novokshonov \cite{AN17,AN18,No18} implemented the algorithms
of Floyd and Knuth \cite{AN17,AN18,No18}.

Sch\"onhage \cite{Sc79} has proven that allowing subtraction and comparison
increases the power of machines which can only add and do equality iff
allowing division increases the power of such machines.
Simon \cite{Si81} extended these studies, in particular he took register
machines which can add, subtract and compare as the base case and then
looked into details -- beyond what he had done jointly with
Hartmanis \cite{HS74} -- what impact additional operations have.
This work was extended also by other authors as Trahan and
Bhanukumar \cite{TB94}.

The model of Floyd and Knuth \cite{FK90} is indeed well-motivated
from the fact that many of the above investigated relations increase
the computational power of polynomial time operations substantially.
Floyd and Knuth were, however, less interested in comparing their model
with variations than in establishing the power of linear time operations
and a fine-grained time-complexity for natural operations on their model,
here linear time means linear in the number of digits of a binary or
decimal representation of the input numbers (if there are several inputs,
one has to take the sum of the digits needed for each input number plus
the number of the inputs).
Their model is motivated from the idea that a central processing
unit of a computer has only few accumulators or registers which perform
basic arithmetic and other operations and for which few of these registers
are sufficient; when abstracting the model and allowing arbitrarily
large values for the registers, the additive operations turn out to be
an adequate choice, as they, as mentioned before, preserve the
class of polynomial time operations.

The precise model of Floyd and Knuth \cite{FK90} allows the
following primitive operations which count each as $1$ step:
Adding or subtracting two registers and assigning
the value to (a perhaps same or different) third register; conditional
goto in dependence of outcome of the comparison of two registers
with $<,\leq,>,\geq,=,\neq$; unconditional goto;
reading of input into registers and writing of output from registers.
However, Floyd and Knuth \cite{FK90} did not allow operations with constants;
as some programs need to handle constants,
they allowed that one additional input $1$ is read in and
stored in a dedicated register in order to have access to constants.
Floyd and Knuth also considered the domain of all real numbers or
the domain of a proper subgroup of the real numbers other than the integers,
that might explain their reluctance to handle constants.
However, the present work does
not go into details of these data structures different from the
integers and the authors of the
present work think that operations and comparisons with constants are
very natural and should be included into the instruction set.
Furthermore, they consider only the integers $\mathbb Z$ or the
natural numbers ${\mathbb N} = \{z \in {\mathbb Z}: z \geq 0\}$
as domains for the algorithms considered; all algorithms can be done
with integers in linear time, but in some cases, only the subcase of
natural numbers is considered in order to reduce the amount of handling
of trivial cases; for that reason, Floyd and Knuth also pose several
questions for natural numbers rather than integers.

Floyd and Knuth \cite{FK90} were precisely interested in two types
of questions: First, how fast are their algorithms in terms of the order
of steps needed in dependence of input size? Here the precise way of
measuring the size depended on their own algorithms and the question
was mainly whether these can be outperformed. Second, for those
where the time complexity is optimal, what is the number of registers
needed by the addition machine to carry out these operations. Both the
time complexity and the number of registers (as some type of space measure)
are aspects fundamental to computer science. Space has two aspects, (1) the
size of the numbers in the registers and (2) the amount of registers itself.
As (2) is constant, it does not influence the asymptotic space usage;
however, it makes a big difference what this constant is and when too
small, many operations can only be carried out in suboptimal time.
Furthermore, explicit restrictions on (1) might contravene the spirit
of the work which allows the numbers to be unrestricted in size and to
measure the influence of the size only indirectly by its affect on the runtime
of the algorithm, so that (2) is the only real space parameter available.

The expressibility of {\bf CNF-SAT}-formulas depends on the number of literals
allowed per clause; 1 literal per clause restricts expressiveness very much
and allows not to code anything interesting; 2 literals per clause allows
to code so much that counting solutions becomes hard while checking solvability
is still easy; 3 literals per clause makes the problem hard to solve and
{\bf NP}-complete. Similarly, for register machines,
the number of registers available allows
for more and more complex linear time algorithms to be carried out and
it is of scientific interest to find for natural operations like multiplying,
dividing and so on where this threshold is. As the basics were already
known from the works of Hartmanis and Simon \cite{HS74,Si81} as well as others,
the research looked now more at the details. Floyd and Knuth were able to
determine optimally the computation time and the number of needed registers
for the greatest common divisor of two numbers
(linear time, 3 registers) and obtained for other operations like multiplying
and dividing the optimal time bound while they were unsure of the number
of registers needed (Question (2)). Questions (3), (4) and (5) then are
questions where they obtained a good algorithm, but could not prove its
optimality with respect to computation time; for these topics the number
of registers were secondary to them, though they are important.
Here is the precise list of the open questions of Floyd and Knuth \cite{FK90}:
\begin{quote}
\begin{enumerate}[(1)]
\item Can the upper bound in Theorem 1 (in \cite{FK90})
      be replaced by $8 \log_\Phi N+\beta$?
\item Can an integer addition machine with only $5$ registers compute $x^2$ in
      $O(\log x)$ operations? Can it compute the quotient $\lfloor y/z\rfloor$
      in $O(\log y/z)$ operations?
\item Can an integer addition machine compute $x^y$ mod $z$
      in $o((\log y)(\log z))$ operations, given $0 \leq x,y < z$?
\item Can an integer addition machine sort an arbitrary sequence of
      positive integers
      $\langle q_1,q_2,\ldots,q_m\rangle$ in
      $o((m+\log(q_1 \cdot q_2 \cdot \ldots \cdot q_m) \log m)$ steps?
\item Can the powers of $2$ in the binary representation of $x$ be computed
      and output by an integer addition machine in $o((\log x)^2)$ steps?
      For example,
      if $x=13$, the program should output the numbers $8,4,1$ in some order.
      (This means, it does not need to be the top-down default order.)
\item Is there an efficient algorithm to determine whether a given $r \times r$
      matrix of integers is representable as a product of matrices of the form
      $I+E_{i,j}$? ($I$ is the diagonal matrix of the identity mapping and
      $E_{i,j}$ has a single $1$ and everything else $0$ at the coordinate
      $(i,j)$.)
\end{enumerate}
\end{quote}
For details of \cite[Theorem 1]{FK90} and further explanations to the
notions of (1) and (6), please consult their paper; the present paper
addresses only questions (2)--(5).
In particular, the present work provides positive answers to (2) and (5):
Floyd and Knuth \cite{FK90} showed that one can compute the greatest common
divisor with three registers in linear time
and they also showed that, in the absence of constants
as operands, this number is optimal. For the operations listed in (2)
they need six registers. Theorems~\ref{thm:four} and~\ref{thm:five}
below show that one can solve the operations in (2) with four registers
used --- or, if one like Floyd and Knuth \cite{FK90} does not allow operations
with constants, the algorithm needs five registers and still matches
the bound of the Open Question.
The operations considered in Problem (2) are to compute the square
of a number and the integer division in the time needed by the algorithms
of Floyd and Knuth which used six registers, but to bring down the register
number to at most five. The runtime for the squaring has to be $\Omega(n)$
where $n$ is the number of bits of the input while for the integer division,
the algorithm has to run in time linear in the number of bits of the output,
note that the latter can be smaller than the number of bits of each
of the inputs.

In Question (5),
Floyd and Knuth \cite{FK90} looked at the binary representation
of natural numbers and wanted
to know the powers of two involved. In other words, given an unknown finite
set $E \subseteq \mathbb N$ and reading the input $x = \sum_{y \in E} 2^y$,
can an addition machine compute and list out all terms $2^y$ with $y \in E$
in any order in time $o(n^2)$? The answer is affirmative and the basic
algorithm is to translate a number given as binary
input $b_1 b_2 \ldots b_n$ by reading the bits $b_1,b_2,\ldots$ at the
top position and adding up the corresponding powers so that one gets
the binary number in reverse order $b_n b_{n-1} \ldots b_2 b_1$
and then to read out the bits at the top while doubling up a variable from
$1$, $2$, $4$, $\ldots$ and outputting $2^{n-m}$ in the case that $b_m = 1$.
This algorithm runs in time $O(n)$ and since in some cases $n$ numbers are
output, the runtime is optimal.

For (3) and (4), answering these questions depends on 
what the expression $o(f(m,n))$ for multivariate Little 
Oh Calculus precisely
means. There are various competing definitions and the followings are
the two popular notions where the first is taken from
the Wikipedia page of the Big Oh Calculus:
\begin{quote}
\begin{enumerate}[(a)]
\item The definition on Wikipedia, based on the algorithms textbook
      of Cormen, Leiserson, Rivest and Stein \cite[Exercise 3.1-8]{CLRS09}:
      A function $f(m,n)$ is in $O(g(m,n))$ iff there exist a constant
      $c>0$ and numbers $m_0,n_0$ such that whenever $m \geq m_0$ or
      $n \geq n_0$ then $f(m,n) \leq c \cdot g(m,n)$.
      Due to this ``or'', it makes according to the Wikipedia page
      a difference whether the natural numbers start with $0$ or with $1$:
      $m+n \in O(m \cdot n)$ if the variables start with $1$
      while this fails in the case
      that variables start with $0$ (choose $m=0$ and 
      let $n=0,1,2,3,\ldots$). 
      Cormen, Leiserson, Rivest and Stein ask readers to form the
      other concepts analogously. So for multi-variate Little Oh expressions
      one can define that $g(m,n) \in o(f(m,n))$
      iff for every rational $c > 0$ there is a $c' > 0$ such that
      for all $(m,n)$ with $\max\{m,n\} \geq c'$, $g(m,n) \leq c \cdot f(m,n)$.
\item Another popular definition of multivariate Big Oh and Little
      Oh Calculus does not use a maximum like Wikipedia but a minimum.
      For the Little Oh notation, this means that $g(m,n) \in o(f(m,n))$
      iff for every rational $c>0$ there is a constant $c'$ such that
      whenever $c' \leq \min\{m,n\}$ then $g(m,n) \leq c \cdot f(m,n)$.
\end{enumerate}
\end{quote}
One can define variants (a) and (b) analogously if more than two variables
are involved or, as in (4), the number of variables is a variable itself.
Note as an example the following difference:
$m+n \in o(m \cdot n)$ for (b) but not for (a), independently on where the
natural numbers start (as long as the start-value is a constant).
Howell \cite{Ho08,Ho12} calls the above $O_\exists$ and $O_\forall$ and
argues that both do not generalise fundamental properties of the one-variable
case and therefore he provides his own third definition.

The findings are now the following: When one takes the Wikipedia definition
(a) of the Little Oh Calculus, one can answer both (3) and (4) to the negative,
that is, these algorithmic improvements do not exist. However, consider
a special case of (4) where a register machine first reads $m$, then
$q_1,q_2,\ldots,q_m$ and then a $k$ and then has to output $q_k$. In this
special case the register machine has only to archive, but not sort the
numbers. Now for Definition (a) of the Little Oh Calculus,
the answer is that this cannot be done in
$o((m+\log(q_1 \cdot q_2 \cdot \ldots \cdot q_m) \log m)$ steps.
However, for version (b) of the definition, the answer would
be that it is indeed possible (where now the minimum of
$m$ and all the read values of $q_1,\ldots,q_m$ have to be above $c'$).
Thus the answer to questions (3) and (4) might indeed be quite sensitive
to what underlying definition for the multivariate Oh Calculus is chosen
and therefore these questions are only partially answered until answers
for all common variants of the Little Oh Calculus are found.

The work on this paper revealed a close connection between automatic
functions and the number of registers in a register machine;
here an automatic function is a generalisation of the notion
of regular sets to that of functions, for more details on this
topic, see Section~\ref{sec:autofunc} below.
For machines having one register only, all functions computed
are, perhaps partial, automatic functions, independently of the
computation time used; furthermore, they are a quite restricted
subset of the set of all automatic functions. On the other hand,
automatic functions can all be implemented with only few registers.
For automatic functions with several inputs (this number of inputs
has to be constant), one can compute them with five registers.
Furthermore, registers machines with two registers can compute
nonautomatic functions in linear time; even in the setting of Floyd
and Knuth where there are no operations with constants.

The results in this paper were jointly developed by the authors
when Xiaodong Jia wrote his UROP thesis about this topic \cite{Jia20}.

\section{On the Methods and Notions in this Paper} \label{sec:methods}

\begin{remark}[Allowed Commands]
In the following let $x,y,z$ be registers (which might refer to the same
register in this definition only) and
$k$ be an integer constant and let $\ell$ be a line number (which
is constant in each case and cannot be varied, for variations in jump
targets, one uses if-then-else commands).
Furthermore, $R$ ranges over the comparison operators $<,=,>,\neq,\leq,\geq$.
The following type of commands are allowed:
\begin{enumerate}
\item Let $x=y+z$; let $x=y+k$; let $x=y-z$; let $x=y-k$; let $x=k-y$;
      let $x=k$;
\item Read $x$; Write $x$;
\item If $x\,R\,y$ then begin $\ldots$ end else begin $\ldots$ end;
\item If $x\,R\,k$ then begin $\ldots$ end else begin $\ldots$ end;
\item Goto $\ell$.
\end{enumerate}
The else-part of if-then-else statements might also be omitted;
similarly bracketing by ``begin'' and ``end'' might be omitted
for single statements.
Below additional constructs will be allowed, as long as
those can be translated into the above constructs in a way that
the number of operations only increases by a constant per use
of the construct and that the number of registers used in total
by the program is not changed. These additional constructs therefore
increase only the readability without doing an essential change.

Read and write commands count also as unit commands and read or write
a full register content. The size of the input can be arbitrary, but
the size parameter $n$ of a function depends on the size of all inputs
read. If it is one input then the size is just the number of binary bits,
that is, the least number $n \geq 1$ such that $-2^n < x < 2^n$. This number
is a rounded version of the base $2$ logarithm of the input number and
its precise definition is the number of binary digits needed to write down
the full number.
Note that $O(\log(x))$ is always the same as $O(n)$ for the number $n$
of digits of an $x>0$ written using an arbitrary but fixed base $b \geq 2$.
For several inputs, the corresponding theorem will always say how
the input size is measured; also please refer to the discussion in the
Introduction on the Big Oh calculus involving several variables.
\end{remark}

\begin{remark}[Usage of Constants]
Floyd and Knuth \cite{FK90} allowed the usage of integer constants only
indirectly by reading the constant once from the input and storing it
in some register where it is available until no longer needed. The main
constant needed is $1$ ($0$ is the difference of a register with itself).
However, this one additional register is also
enough, as every integer constant to be added or subtracted can be
replaced by adding or subtracting the $1$ a constant number of times.
Similarly for comparing $x$ with $k$, one subtracts $k-1$ from $x$,
does the comparison and adds $k-1$ back to $x$.

Though operations with constants are an obstacle for proving lower
bounds, the authors of this paper think that allowing operations
with constants is natural. For example, early CPUs like MC6800 from
Motorola could add and subtract and compare either one register
(called accumulator) with another one or with constants, furthermore,
the CPU could not multiply  -- one of the authors
used a computer with this CPU at his secondary school.

If one wants to translate results on number of registers sufficient for a
computation of a function from the model of this paper to the {\bf model of
Floyd and Knuth}, one has in general to add one to the number of registers
used.
\end{remark}

\begin{remark}[Variables and Registers] \label{rem:three}
When writing programs, one might in addition to the registers also consider
variables which hold values from a constant range, say bits. These
variables do not count for the bound on the number of registers,
as they can be implemented by doubling up the program (in the case
of bits) and then jumping back and forth between the two copies
of the problem, which will then be adjusted to the variable having
the value $0$ in the first copy and the value $1$ in the second
copy; here is an example.
\begin{quote}
\begin{enumerate}
\item Read $x$; read $y$;
\item If $x<y$ then $b=1$ else $b=0$;
\item Let $x=x+y$; let $y=x+x$;
\item Let $y=y+b$;
\item Write $y$.
\end{enumerate}
\end{quote}
An optimised way of implementing this without using $b$ is the
following:
\begin{quote}
\begin{enumerate}
\item Read $x$; read $y$;
\item If $x<y$ then goto 5;
\item Let $x=x+y$; let $y=x+x$;
\item Goto 7;
\item Let $x=x+y$; let $y=x+x$;
\item Let $y=y+1$;
\item Write $y$.
\end{enumerate}
\end{quote}
In the worst case, the full program has to be transformed into
$k$ consecutive copies where $k$ is the number of values the variable can
take. One loads a value into such a variable by jumping into the
corresponding copy; if it is read out, the variable is at each of the
copies of the program replaced by the corresponding integer constant.
Floyd and Knuth \cite{FK90} used a similar method at the program
$P_6$ where they permuted the order of the registers without giving
the code for it.
Letters at the beginning of the English alphabet are used for such constant
range variables, while letters at the end of the English alphabet are used
for registers.
\end{remark}

\section{Solving Open Problems (2) and (5) of Floyd and Knuth}
  \label{sec:twofive}

\noindent
In Open Problem (2),
Floyd and Knuth \cite{FK90} were interested in the optimal number of
registers needed for basic operation on a register machine that can add
and subtract and compare; they considered general subgroups of the reals,
but concentrated on the integers which are also the model of this work.
A side-constraint is that this number of registers should take the time
consumption of the operation into account and so, for multiplication,
remainder and division, it should be in $O(n)$ and not larger.
Floyd and Knuth \cite{FK90} gave an unbeatable algorithm of calculating
the remainder using the Fibonacci numbers to go up and down in exponentially
growing steps from the smaller number to the bigger number and back.
Furthermore, they showed
that the tasks cannot be solved with two registers. While the Fibonacci
method is unbeatable for many items, some of the following problems
withstood attemps to solve them with this method.
The present work addresses the corresponding problems left open in their work.
The next theorem solves the first part of the Open Problem (2) of
Floyd and Knuth \cite{FK90}.

\begin{theorem} \label{thm:four}
Multiplication can be done using four registers in time linear in
the size of the smaller number (in absolute value). In particular
the squaring of a number $x$ can be done with four registers in
linear time, that is, in time proportional to $\log(|x|)$ for $x \geq 2$.
\end{theorem}

\begin{proof}
Multiplication can be done with four registers in time linear in
the size of the smaller number. The algorithm is as follows:
\begin{quote}
\begin{enumerate}
\item Begin Read $x$; read $y$; let $a=1$;
\item If $x < 0$ then begin let $x=-x$; let $a=-a$ end;
\item If $y < 0$ then begin let $y=-y$; let $a=-a$ end;
\item If $y < x$ then begin let $v=x$; let $x=y$; let $y=v$ end;
\item Let $v=1$; let $w=0$; let $x=x+x$; let $x=x+1$;
\item If $v > x$ then goto 7; \\
      let $v=v+v$; goto 6;
\item Let $x=x+x$; if $v =x$ then goto 8;\\
      let $w=w+w$; \\
      if $x>v$ then begin let $w=w+y$; let $x=x-v$ end; \\
      goto $7$;
\item If $a=-1$ then begin let $w=-w$ end; \\
      Write $w$; End.
\end{enumerate}
\end{quote}
The following arguments verify the runtime-properties of the algorithm.
For this let $n$ be the number of binary digits needed to write down
the, by absolute value, smaller one of the numbers $x$ and $y$.
Line 1 reads the input and initialises the sign-variable $a$ as $1$.
Lines 2, 3 and 4 enforce that $x,y$ are updated to the minimum
and the maximum of the absolute values of these two inputs and that
$a$ is the sign from $-1$ and $+1$ with which the nonnegative output
has to be multiplied in the last line 8 before the output; so if either
both $x,y$ were input as negative numbers or none then $a=1$
else $a=-1$. The verification
of these properties is straight forward and omitted.

The idea of the remaining part is to read out the top bit of $x$ and double
$x$ until all bits except for the last one are read out. For this one needs
a perparation in lines 5 and 6: Line 5 sets $v=1$ and $w=0$ and updates
$x$ to $2x+1$. $v$ is the variable which will later hold the least power
of $2$ larger than the just updated $x$. Line 6 doubles $v$ until
$v > x$, thus $v = 10^{n+1}$ in binary after completing the loop in that line
where $n+1$ is the number of the binary digits of the updated $x$, that
is the convention $n=0$ holds in the case that one of the original inputs
is zero.

The second and main loop is in Line 7. Prior to this, the value of
$x$ in binary is $b_n b_{n-1} \ldots b_1 1$ where $b_n b_{n-1} \ldots b_1$
is the smaller absolute value of the two inputs and $b_n = 1$
unless $n=0$.

Now the loop invariant is the following: After $m$ rounds through the
loop in line 7, the value of the variable $w$ is
$b_n b_{n-1} \ldots b_{n-(m-1)}$ times $y$ and the value of $x$ is
$b_{n-m} \ldots b_1 1 0^m$ where $0^m$ means $m$ zeroes.
The initial value of $w$ is thus just $0$ so that the loop invariant
holds initially (after zero times executing the loop).

The loop body does per iteration the following.
First $x$ is updated to $x+x$, that is, takes the value
$b_{n-m} \ldots b_1 1 0^{m+1}$. In the case that $m=n$
this means that $x = v$ and the next conditional goto command to $8$ quits
the loop with $w$ having the value $b_n b_{n-1} \ldots b_1$
times $y$ as required, so that the right output will be given.
Otherwise it is $v \neq x$ and $m<n$. Now $w$ is doubled up so
that $w$ takes the value $b_n b_{n-1} \ldots b_{n-(m-1)} 0$ times $y$.
Now $x \geq v$ iff the top bit $b_{n-m}$ of $x$ is $1$. This is reflected
by the next if-then-statement:
If $x>v$ an update is done that $x=x-v$ (so that the topmost bit of
$x$ is set to $0$ and thus erased) and $w$ is updated to $w+y$.
After that if-command the value of
$x$ is $b_{n-m-1} \ldots b_1 1 0^{m+1}$ and the value of $w$
is $b_n b_{n-1} \ldots b_{n-m}$ times $y$.
Thus the loop invariant from above holds also after doing
the loop body $m+1$ times and the last command in line 7 jumps
to the beginning of the line to do the next iteration of the loop body.

Lines 2,3,8 handle the sign of the operands,
where the product of the sign is stored in $a$ in order to handle the
negative numbers. Remark~\ref{rem:three} explains
that one can have variables which use only constantly many values without
having to increase the number of registers; therefore one does not need
an extra register for variable $a$. Line 4 handles an optimisation
which was not required by Floyd and Knuth and which just orders
$x,y$ as $x \leq y$, not relevant for squaring.

For the runtime, note that there are only two loops, in lines 6 and 7.
The loop in line 6 doubles up a number until it is greater than the input
and this number is initially 1, thus it runs $O(n)$ times where $n$ is
the length of the input in bits. The loop in line 7 runs through the loop
$n+1$ rounds, thus again is $O(n)$ with the parameter $m$ being
$m=0,1,2,\ldots,n$. So the overall runtime is $O(n)$.
\end{proof}

\noindent
This example illustrates the multiplication in
order to make the verification of the algorithm easier.
Now assume that $x$ and $y$ are the binary numbers
$1101$ and $100001$. In line 5, $x$ is updated to 11011 (a coding bit 1
appended), $v$ is initialised to $1$ and $w$ is initialised to $0$.
Now line 6 doubles $v$ until $v>x$, that is until $v$ has one digit more
than $x$ and the values of $x,y,v,w$ are as follows:
\begin{verbatim}
  x     11011
  y    100001
  v    100000
  w         0
\end{verbatim}
In line 7, the registers $x,w$ will be doubled up in each round
with a special exit out of the loop if $x=v$ prior to doubling
$w$ as that means that
the coding bit has reached the position of the $1$ in $v$ and that
the multiplication is complete.
Furthermore, if after doubling up $x>v$ then $v$ is subtracted from $x$
and $w$ is updated to $w+y$.

This is now illustrated by giving the values of the
various numbers during the iterations of the loop using example
from above, note that only $x$ and $w$ change while $y$ and $v$
remain always the same, thus they are only in the first two lines.
\begin{verbatim}
  x     11011 // x before the loop;
  w         0 // w before the loop;
  y    100001 // y throughout the loop;
  v    100000 // v throughout the loop;
  x     10110 // leading 1 read out and v subtracted from x;
  w    100001 // y added to w;
  x     01100 // leading 1 read out and v subtracted from x;
  w   1100011 // w doubled up and y added to w;
  x     11000 // leading 0 read out and no subtraction;
  w  11000110 // w doubled up and no addition;
  x     10000 // leading 1 read out and v subtracted from x;
  w 110101101 // w doubled up and y added to w;
  x    100000 // after the last doubling up of x, v=x and loop end;
  w 110101101 // value of w from last loop body returned as result.
\end{verbatim}
The last line $8$ just multiplies the output with $-1$ to the power of
the number of negative inputs. So in the case that the original input
was either 1101,-100001 or -1101,100001 then the output is -110101101
else it is 110101101.

\medskip
\noindent
Floyd and Knuth \cite{FK90} overcame the obstacle that one cannot
divide by $2$ in constant time in an addition machine by resorting to
the Zeckendorf representation of natural numbers
where each natural number is the unique sum of
non-adjecent Fibonacci numbers \cite{Ze72}.
When multiplying $v$ with $x$, they had
auxiliary register $u,w,y,z$ where initially one reads $v,x$
and sets $w = v$ and $u=x-x$ and $y = z = 1$.
Part of the invariant is that $y \leq z$, $y,z$ are adjecent Fibonacci numbers,
$v = y \cdot v'$ and $w = z \cdot v'$ where $v'$ is the initially read
value of $v$. After initialisation, the constant $1$
is not needed again. So $y,z$ are the first Fibonacci numbers and on
the way up, one does the following updates:
\begin{quote}
  (1) $y = y+z$; $v = w+v$; swap $y,z$; swap $v,w$;
\end{quote}
until $y \leq x < z$. Note that swapping can be obtained
by renaming the variables without doing any real operation.
Once this is achieved, one updates
\begin{quote}
   (2) $x = x-y$; $u = u+v$;
\end{quote}
and then does repeatedly the updates
\begin{quote}
   (3) $z = z-y$; $w = w-v$; swap $y,z$; swap $v,w$;
\end{quote}
until either $x< y = z$ (that is, $x=0$ and $y=1$ and $z=1$)
or $y \leq x < z$. In the first case the program
terminates with output $u$, in the second case it goes to statement (2)
and continues from there.

Floyd and Knuth \cite{FK90} resorted to the Zeckendorf representation
of numbers for the reason that one
can go up and go down in this representation; however, one needs two
registers to store two neighbouring Fibonacci numbers, namely $y,z$;
furthermore, one needs $v,w$ to store the products of $y \cdot v'$ and
$z \cdot v'$, concurrently, as well, where $v'$ is the initial value
of $v$. So on one hand, due to the bidirectionality, the Fibonacci
representation is easier to use than powers of $2$ for the same purpose,
on the other hand, one needs instead of one register to store a power
of $2$ now two registers to store two neighbouring Fibonacci numbers
as well as two registers to store the product of one of the factors
with these neighbouring Fibonacci numbers. Thus this algorithm, though
more flexible with respect to going down in the Fibonacci numbers,
needs more registers than the algorithm of Theorem~\ref{thm:four}.
The obstacle of generating the powers of two summing up to an input
number in linear time was not resolved by Floyd and Knuth, but left as
Open Problem (5).
Thus their approach of using the Zeckendorf representation was their way
around this obstacle. A part of the algorithm to multiply binary numbers
is to identify the corresponding powers of two which occur in a factor,
thus Open Question (5) is asked in order to decide whether one can at
least do the easier part of the multiplication, that is, to extract the
powers of two which sum up to a given factor in linear (or even subquadratic)
time.

The next theorem solves the second part of Open Problem (2) of
Floyd and Knuth \cite{FK90}.

\begin{theorem} \label{thm:five}
The integer division $x = Floor(y/z)$ with $z \neq 0$
can be carried out with 4 registers
in time $O(m)$ where $m$ is the smallest natural number such that
$2^m$ times the absolute value of $z$ is greater or equal the absolute
value of $y$.
\end{theorem}

\begin{proof}
There are four registers, $x,y,z$ are related to the input and output
of the algorithm as specified in the theorem and $u$ is the least number
of the form $z \cdot 2^\ell$ which is at least as large as $y$
(after $y,z$ have been made positive to avoid sign-problems).
As in the algorithm for Theorem~\ref{thm:four},
the key idea is to read out binary numbers
at the top by comparing them with a power of $2$; here however, one
compares not with $2^\ell$ but $2^\ell \cdot z$ and the termination
condition is that $z$, which will be doubled up as well, reaches $u$.
Besides initialising $x=0$ and $u=z$, the main function of lines $1$,
$2$, $5$ and $6$ is to read the inputs $y,z$, to handle the sign
and to write the output $x$; lines $3$ and $4$ compute integer division
for $y \geq 0$ and $z \geq 1$ and the reader can concentrate on this
part to understand the key ideas of the algorithm.
The formal algorithm is as follows.
\begin{quote}
\begin{enumerate}
\item Begin read $y$; read $z$; let $x=0$;
\item If $z<0$ then begin $z=-z$; $y=-y$ end; \\ if $z=0$ then goto 6; \\
      if $y<0$ then begin $a=-1$; $y=-y$ end else begin $a=1$ end; \\
      let $u=z$;
\item If $u \geq y$ then goto 4; $u=u+u$; goto 3;
\item if $y \geq u$ then begin $y=y-u$; $x=x+1$ end; \\
      if $u = z$ then goto 5; \\
      let $x=x+x$; let $y=y+y$; let $z=z+z$; goto 4;
\item If $a=-1$ then begin let $x=-x$; if $y>0$ then let $x=x-1$ end;
\item Write $x$ end.
\end{enumerate}
\end{quote}
The next paragraphs give the verification for the main case that
$y \geq 0$ and $z \geq 1$. The other cases are left to the reader;
due to these main assumptions, the instructions in Line 2 (except for letting
$u=z$) and Line 5 can be ignored, as they handle the exceptions for
the case that the above assumption is not satisfied. The output
in the case of division by $0$ is irrelevant.

If $0 \leq y < z$ then the loop in Line 3 is skipped and the
if-then statement at the beginning of line 4 leaves $x,y$
unchanged and the next statement has $u=z$ (due to the loop in line 3
not being done) and returns $0$, this shows correctness for
this basic case.

While register $u$ is not greater or equal to $y$, $u$ is doubled
up in Line 3. Note that therefore $u = 2^m \cdot z$ for the
minimal $m$ with $2^m \cdot z \geq y$ after processing Line 3;
$m$ can be $0$ and $m$ is identical with the above same-name parameter
of the runtime of the algorithm.

For Line 4, let $y',z'$ be the respective values of $y,z$ before
entering the line which are the absolute values read of what
has been read into $y,z$ at the beginning of the program.

The invariants of Line 4 is that after $s$ rounds of this line,
$y$ has the value $2^s \cdot y' - x \cdot z' \cdot 2^m$
and $z$ has the value $2^s \cdot z'$ and $0 \leq y < 2u$. In round $s$,
the algorithm first checks whether $y \geq u$ and if so,
subtracts $u$ from $y$ and increments $x$ by $1$, in other words,
$2^m \cdot z'$ gets subtracted from $y$ and $1$ added to $x$
so that the difference $y-x \cdot z' \cdot 2^m$ before and after
this update are the same; furthermore, after the update the property
$0 \leq y-x \cdot z' \cdot 2^m < u$ holds.
If now $u = z$ then the algorithm quits the loop
else it doubles up $x,y,z$ and goes into the next round of the loop.
The algorithm indeed quits after $m$ rounds, as $z$ is initially
$z'$, doubled up exactly once in each round and $u = 2^m \cdot z'$.
When the algorithm quits the loop,
then $0 \leq 2^m \cdot y' - x \cdot 2^m \cdot z' < u = 2^m \cdot z'$ and,
when dividing by $2^m$, one sees that
$0 \leq y' - x \cdot z' < z'$ so that $y-x \cdot z'$ is indeed
in $\{0,1,\ldots,z'-1\}$ as the integer division by $z'$
requires. Thus $x$ is the intended downrounded value of $y'/z'$.

The loops in lines 3 and 4 run $m$ times, as in line 3 one measures
how often one has to double up $u$ (which was $z$) to reach or overshoot
the value of $y$ and the loop in line 4 brings up $z$ by doubling up
this register until $u$ is reached. The handling of the sign is done
in lines 2 and 5 and the values of $x,y,z,u$ are at least $0$
in the loops in 3 and 4.
\end{proof}

\noindent
In Open Problem (5), Floyd and Knuth \cite{FK90} ask whether
there is a register machine which can in subquadratic time compute the powers
of $2$ giving the sum of a given number $x$ in arbitrary order
(but each outputting only once). For example, for $x=100$, the algorithm
should output $4,32,64$ in arbitrary order.
The next algorithm for this runs in linear time and needs
only four registers; thus the algorithm satisfies the subquadratic
runtime bound requested by Floyd and Knuth \cite{FK90}.

\begin{theorem} \label{thm:six}
On input of a number $x \geq 0$ of $n$ digits, a register machine with
four registers can output the powers of two giving the sum $x$ in time linear
in $n$.
\end{theorem}

\begin{proof}
The idea is to first reverse the bits in the representation of $x$
and then to read out the powers from the top,
now using that the $k$-th bit stands for $2^k$.
\begin{quote}
\begin{enumerate}
\item Begin read $x$; if $x<0$ then begin let $x=-x$ end; \\
      let $y=1$; let $z=1$; let $x=x+x;$ let $x=x+1$; let $u=0$;
\item If $y > x$ then goto 3; let $y=y+y$; goto 2;
\item If $y=x$ then goto 4; \\
      if $x > y$ then begin let $u=u+z$; let $x=x-y$ end; \\
      let $x=x+x$; let $z=z+z$; goto 3;
\item Let $z=1$; let $x=u+u$;
\item If $x \geq y$ then begin write $z$; let $x=x-y$ end; \\
      let $z=z+z$; let $x=x+x$; if $x > 0$ then goto 5; \\
      End.
\end{enumerate}
\end{quote}
First one makes sure that $x$ is not negative. Then
one enters into $x$ a termination condition which is an additional
$1$ at the end in order to put the correct number of zeroes when
inverting the number. This is all done in line 1.

The first loop in line 2 determines a power of $2$ which is a proper
upper bound on $x$. This bit of this bound is two digits ahead of the
largest power of $2$ in the sum of $x$. So $x$ holds the original input
$b_1 b_2 \ldots b_n$ appended with a coding bit $1$, where $b_1 = 1$.
Furthermore, the register $y$ holds the number $2^{n+1}$.

The next loop in line 3 converts the number $b_1 b_2 \ldots b_n 1$
to $1$ times $2^{n+1}$ in $x$ and $b_n b_{n-1} \ldots b_1 0$ in $u$;
for this one starts with $u=0$ and $z=1$ and the loop invariant of
this loop is that after $m$ rounds (with $m \geq 1$) of the loop,
$x$ has the form $b_m b_{m+1} \ldots b_n 1$ times $2^m$ and $z$ has the
value $2^m$ and $u$ has the value $b_{m-1} b_{m-2} \ldots b_2 b_1 0$.

In each round of the loop, first it is checked whether $x=y$ and
if so, the loop is left. Then it is checked whether $x>y$ and if
so, $y$ is subtracted from $x$ and $z$ is added to $u$. At the
end, $x$ and $z$ are doubled up. In the first round, only the
doubling up happens, as $x<y$. After $m$ rounds with $m \geq 1$,
$x$ has the value $b_m b_{m+1} \ldots b_n 1$ times $2^m$,
$z$ has the value $2^m$ and $u$ has the value
$b_{m-1} b_{m-2} \ldots b_1 0$.

In line 5 the powers of two are output whenever the leading bit of
$x$ is $1$ and the leading bit is removed from $x$, after that $x$
and $z$ are doubled up; when $x=0$ the loop terminates.
So the invariant is that after $m$ rounds of the loop in line 5,
$x$ is $b_{n-m} b_{n-m-1} \ldots b_1$ times $2^{m+1}$, $2^{m-1}$
has been output iff $b_{n-m+1}$ was $1$ and $z$ is $2^m$.
As the number $x$ was in binary at the input $b_1 b_2 \ldots b_n$
or $- b_1 b_2 \ldots b_n$,
outputting $2^{m-1}$ iff $b_{n-m+1} = 1$ is correct.

Each of the three loops runs approximately $n$ times where $n$ is the
number of binary bits in $x$, thus the runtime is $O(n)$.
Note that in each loop, $y$ and $x$, respectively, are doubled up
until the termination condition is reached. In line 2, the termination
condition is just $y>x$ what will be reached as $y=1$ initially.
In line 3, the termination condition is $y=x$ what will be reached,
as $y = 2^{n+1}$ and in each round, the bit in $x$ at the position of the
$1$ in $y$ is removed and the remaining number doubled up until $y=x$;
the latter condition is reached as the last bit of $x$ as at the beginning
of step 3 is
a $1$. In Line 5, again the bit of $x$ at the position of the $1$ in $y$
is read and removed in each round and $x$ is doubled up
until the loop terminates when $x=0$. Thus all three loops run
$O(n)$ times and the overall runtime of the algorithm is $O(n)$.
\end{proof}

\noindent
For example, one wants to output the powers of two in the number thirteen,
in binary $1101$. In line $1$, the
number gets transformed into $11011$ by appending a bit.
Then $y$ is set to $1$ and doubled in line $2$
until it is greater than $x$, that is, takes the value $100000$.
In line $3$ the algorithm runs through five rounds and the
values of $y,x,z,u$ after rounds $1,2,3,4,5$ are as follows:
\begin{verbatim}
  y 100000 // remains like this through-out the loop
  x 110110 // coding bit appended and doubled up in round 1 
  u      0 // initialised as 0 and not modified in round 1
  z     10 // doubled up in round 1
  x 101100 // bit 1 read out and doubled up in round 2
  u     10 // old z added to u in round 2
  z    100 // doubled up in round 2
  x 011000 // bit 1 read out and doubled up in round 3
  u    110 // old z added to u in round 3
  z   1000 // doubled up in round 3
  x 110000 // bit 0 read out and doubled up in round 4
  u   0110 // nothing added in round 3, leading zero for readability
  z  10000 // doubled up in round 4
  x 100000 // bit 1 read out and doubled up in round 5
  u  10110 // old z added to u in round 5
  z 100000 // doubled up in round 5
  x 100000 // not modified as x=y in round 6, loop terminated
  z      1 // z is set to 1 in line 4 after termination of loop
  x 101100 // x is doubled up in line 4
\end{verbatim}
Here the values in the loop in line 5  with the above data:
\begin{verbatim}
  y 100000 the register y remains unchanged
  x 101100 before round 1
  z      1 before round 1
  x 011000 bit 1 is read out, y subtracted from x and x doubled up in round 1
         1 is output in round 1
  z     10 z is doubled up after outputting z in round 1
  x 110000 bit 0 is read out and x is doubled up in round 2
  z    100 z is doubled up without being output in round 2
  x 100000 bit 1 is read out, y subtracted from x and x doubled up in round 3
       100 is output in round 3
  z   1000 z is doubled up after outputting z in round 3
  x 000000 bit 1 is read out, y subtracted from x and x doubled up in round 4
      1000 is output in round 4
     10000 z is doubled up after outputting z in round 4
  x 000000 loop terminates after round 4 as x = 0
\end{verbatim}
%The above arguments verify the correctness.
%Each of the three loops runs approximately $n$ times where $n$ is the
%number of binary bits in $x$, thus the runtime is $O(n)$.
%Sanjay: removed above lines as the above is just example of
%the theorem. Proof of theorem is earlier.

\section{Open Problems (3) and (4) and Multivariate Oh Notations} \label{sec:threefour}

\noindent
There are several ways to define the Big and Little Oh Notations
in several variables. Wikipedia (version (a)) gives with reference to
Cormen, Leiserson, Rivest and Stein
\cite[page 53]{CLRS09} for the following definition:
$f(x_1,x_2,\ldots,x_k)$ is in $O(g(x_1,x_2,\ldots,x_k))$
if there are constants $c,d$ such that for all tuples
$(x_1,x_2,\ldots,x_k)$ where at least one coordinate is above $d$,
$f(x_1,x_2,\ldots,x_k) \leq c \cdot g(x_1,x_2,\ldots,x_k)$.
The analogous definition for $f \in o(g(x_1,x_2,\ldots,x_k))$ is
that for all constants $c>0$ there is a constant $d$ such that
for all tuples with $(x_1,x_2,\ldots,x_k)$ with at least one
of the coordinates above $d$,
$f(x_1,x_2,\ldots,x_k) \leq c \cdot g(x_1,x_2,\ldots,x_k)$.
If one would not require that only one coordinate is above $d$
but all coordinates are above $d$, the next result is not applicable.
Version (b) of the multivariate Little Oh Calculus requires that
not only one coordinate but all coordinates are above $d$ and,
in the case that the number of coordinates varies as well, that
there are at least $d$ coordinates in the tuples considered.

Floyd and Knuth \cite{FK90} asked whether one can compute $x^y$ modulo $z$
in time $o(n \cdot m)$ where $n$ is the number of digits of $y$ and $m$
is the number of digits of $z$. The following example answers Question
(3) only for the Wikipedia definition (variant (a)) of the Little Oh Calculus

\begin{exmp} \rm
In this example, when denoting values modulo $z$, in order to
estimate their size modulo $z$, these are numbered as
$-z/2,-z/2+1,\ldots,-1,0,1,\ldots,z/2-1$ and not as $0,1,\ldots,z-1$,
as part of this example requires to study the first nonzero bit of such
numbers. $z$ is always even.

One chooses $y>2$ 
%arbitrarily 
and $z$ to be so large that the constant of
the little $o$ is below $0.1/(y \cdot \log(y))$. Furthermore, $x$ is
$2^{m/(y+1)}$ and it is understood that $z$ is chosen such that $x > y$
and that $x$ is an integer.
Now one estimates that at every operation (addition or subtraction)
of the register machine, the largest register increases its value,
modulo $z$, by at most a factor $2$.
Note that, modulo $z$, the largest input is $x = 2^{m/(y+1)}$ and
that the output is $x^y = 2^{m \cdot y/(y+1)}$ which is smaller than $z/2$.
Thus one would need that, modulo $z$,
the first nonzero digit of the largest registers goes from
$m/(y+1)$ to $m  \cdot  y/(y+1)$ which requires at least $m \cdot (y-1)/(y+1)$
additions.
This amount of additions is larger than $c  \cdot  m  \cdot  n$,
as $c \leq 0.1/(n  \cdot  y)$ and
$(y-1)/(y+1) \geq 0.5$, so the algorithm cannot make enough additions
and subtractions for producing a result which, modulo $z$, equals $x^y$.
\end{exmp}

\begin{theorem} \label{thm:problemfour}
Consider the task that a register program
reads in a positive number $m$ followed by $m$ positive
numbers $q_1,\ldots,q_m$ followed by one number $k \in \{1,2,\ldots,m\}$
in this order and has then to output $q_k$. In the following,
let $r = \log(\max\{2,q_1,q_2,\ldots,q_m\})$ and let
$f(m,r) = r \cdot m \cdot \log(m)$.
This task cannot be done in $o(f(m,r))$ steps in the case that one
applies Definition (a) of the Little Oh Calculus and it can be done
in $o(f(m,r))$ steps in the case one applies Definition (b) of the
Little Oh Calculus.
\end{theorem}

\begin{proof}
For the result concerning Definition (a) of the Little Oh Calculus,
given a register program with $s$ registers, let $m = 2s$ and fix
it at this constant and let
$r$ be so large that all tuples of $m$ $r$-digit numbers
the input has to be processed in time $f(m,r) / (m^2 \log(m)) = r/m$
- what is possible as $m$ is now fixed in the Little Oh Calculus
and the $r$ is chosen so large that the runtime is smaller than
$f(m,r)$ times the rational number $1/(m^2 \log m)$.
At the same time, as $2m$ $r$-bit numbers are read,
the machine must save them in its registers and be able to recall
each of them and also know the position of each number. Thus there
are after reading the $m$ numbers $2^{m (r-1)}$ many different
$m$-tuples of $r$-bit numbers (with leading bit $1$ at the top position).
It is now not possible to store them in a one-one way in $s$ registers
if all $s$ registers have numbers strictly below $2^{2r-3}$, as those
jointly use only $m/2 \cdot (2r-3) = m(r-1.5)$ bits and can take
only $2^{m(r-1.5)}$ many values. If two $m$-tuples are mapped to the
same memory and differ on item $k$, then the algorithm will for one
of the $m$-tuples make a mistake when the next number read is $k$.
Thus one of the numbers must at least have $2r-2$ bits. However,
when reading only $r$-bit numbers and the smaller value of $m$,
there must be at least $r-2$ additions or subtractions in order
to create a number which has properly $2r-2$ bits (with the
highest order bit being $1$). As $m \geq 2$ and
$r-2 > r/m$ for all sufficiently large $r$, the computing
task is not in $o(f(m,r))$ when the Little Oh Calculus is taken
according to Definition (a).

For Definition (b), the idea is to prove that the task is in
$O(m \cdot r)$. By the definition of (b), $O(m \cdot r) \subseteq
o(f(m,r))$ as $lim_{m,r \rightarrow \infty} m \cdot r/f(m,r) =
\lim_{m,r \rightarrow \infty} 1/\log(m) = 0$ provided that both
$m,r$ go to infinity and not only their maximum (as (a) requires).

The algorithm is to use $O(m \cdot r)$ operations to create a queue
which is fed at the bottom and read out at the top. For the ease of
readability of the program, all numbers in the input are required
to be positive (so at least $1$) and this is not tested explicitly
(though it would be trivial to do so).
\begin{quote}
\begin{enumerate}
\item Read $v$; read $w$; let $u=1$; let $x=0$; let $y=0$; let $z=1$;
\item let $x=x+x$; let $y=y+y$; let $z=z+z$; $u=u+u$; \\
      if $z \leq w$ then goto 2;
\item Let $v=v-1$; let $x=x+w$; let $y=y+1$; let $z=1$; \\
      if $v<1$ then goto 4 else begin Read $w$; goto 2 end;
\item Read $v$; let $z=0$;
\item Let $x=x+x$; $y=y+y$; let $z=z+z$; \\
      If $u \leq x$ then begin let $x=x-u$;
        if $v=1$ then let $z=z+1$ end; \\
      if $u \leq y$ then begin let $y=y-u$; let $v=v-1$ end; \\
      if $(y>0$ and $v>0)$ then goto 5;
\item Write $z$.
\end{enumerate}
\end{quote}
This program produces a data structure where the number $u$ determines
the top position of the data structure and $x$ has the bits of the numbers
one after the other and $y$ has the end positions of each binary number
in the structure. So when entering the if-then-else statement at the end
of line 3, the data structures for the so far processed
binary numbers $110$, $101$, $11011$ looks like this:
\begin{verbatim}
  u  1 000 000 00000
  x    110 101 11011
  y      1 001 00001
  z                1
\end{verbatim}
Furthermore, $v$ contains the remaining numbers to be built into the
data structure and $w$ contains the most recent number $11011$ added
into the data structure.

So when building up the data structure, the role of $z$ is to space
out the numbers so that when adding $w$ to $x$, the bits will not overlap
with those of the previous number and therefore $u,x,y,z$ are doubled up until
$z > w$. Furthermore, when $w$ is added to $x$, $1$ is added to $y$ in
order to mark the last bit in each round. The inner loop of doubling
up is in line 2 and the outer loop also includes line 3 to do the additions
of the current number $w$ to $x$ and of the current end-bit marker to $y$.

In the loop of Line 5, the bits of $x,y$ are read out in parallel
by always doubling up so that the position of the leading bit (it might
be $0$) is moved at the position of the only $1$ of $u$ in binary
representation and then one makes two if-statements one dealing with
$x$ having a $1$ in this leading position and one dealing with $y$
having a $1$ in this leading position. These leading digit of $x$ is
copied into the last position of $z$ and that of $y$ causes, when being
$1$, the counter $v$ to be decreased. If $v$ is still positive, that is,
if $z$ is still $0$, and if $v$
is decremented to $0$ it means that the number currently in $z$ is the
number to be passed into the output. If the number $v$ was too big (and
there is no number archived for that index) then $y$ will eventually become
$0$ and the loop will be aborted and some meaningless output be given.
Here the above examples after two bits of $x,y$ are read out in the case
that the value of $v$ is $1$ (what causes the bits of $x$ to be copied
to $z$):
\begin{verbatim}
  u  1 0 000 00000 00
  x    0 101 11011 00
  y    1 001 00001 00
  z                11
\end{verbatim}
The $11$ in $z$ are the two first bits of the binary number $110$ which
was coded as first number in $x$. The last bit, a $0$, is still in and
written here as a leading bit for better readability. The spaces in the
number are for readability and shifted to the front inline with the
doubling up of the numbers, only $u$ remained unmodified and there the
spaces are just adjusted to those of $x$ and $y$ for having them at the
same positions.

This algorithm verifies that the task of the problem is in
$O(m \cdot r)$ as first the numbers $x,y,u$ which are doubled up
in every round of the loop will have at the end $m$ numbers of
up to $r$ bits -- more precisely $r_1+r_2+\ldots+r_m$ bits in the
case that the $k$-th input into the register $w$ has $r_k$ bits.
Here note that $m$ is the first input into $v$ and the second input
into $v$ is the index of the number to be read out (from the front).
The second loop also runs at most the same number of rounds, as after
this number of rounds the value of $y$ is $0$. $y$ and $x$ get doubled
up in each round and their top bits, which are at the position of the
single bit one of $u$, are removed by subtracting if they are not zero.
For that reason, the runtime of the algorithm is
$O(r_1+r_2+\ldots+r_m)$ where each $r_k \geq 1$. This is upper bounded
by $O(r \cdot m)$ and, for the notion (b) of the Little Oh Calculus,
$O(r \cdot m) \subseteq o(r \cdot m \cdot \log(m))$. Note there that
for a constant $c' > 0$, one chooses $m$ so large that $\log(m) \cdot c'$
is above the multiplicative constant of the runtime expression $O(r \cdot m)$
and therefore the runtime is less or equal $c' \cdot r \cdot m \cdot \log(m)$.
\end{proof}

\begin{remark}
The above implies that Open Problem (4) is answered ``no'' when one
bases the Little Oh Calculus on version (a) which is the one on
Wikipedia, as the task in Open Problem (4) is more comprehensive than
the one in the preceding theorem.
%, which is a significantly easier
%task than that of Problem (4) and this task also does not run in
%the given time bound provided version (a) of the Little Oh Calculus
%is chosen.  
Furthermore, the result shows that there is a real chance
that the answer of problems (3) and (4) might actually depend on the
version of the Little Oh Calculus chosen, so it could go in either 
direction. But this ``actual chance'' is not yet converted
into a proof, but only indicated as a possibility. On one hand, disproving
the existence of an algorithm for version (b) is much harder than for
version (a) and on the other hand, algorithms
confirming that the answer would be ``yes'' (as in the case of Problem (2))
are not in sight.

The proof of Theorem~\ref{thm:problemfour} also
showed that, for constant $m$, a register machine
with $m$ registers needs steps proportional to the number of bits
of the input numbers to recall one out of
$2m$ read inputs which have all the same length; in contrast to this,
the machine with $m$ registers can recall any of $m-1$ inputs
by storing them in the first $m-1$ registers and using the
$m$-th to read the index of the recalled number.
Similarly for sorting constantly many numbers, the time depends
on the length of the numbers only in the case that the number
of registers is below the number of inputs. This shows
that when considering the asymptotically fastest machines only,
the number of registers forms a proper hierarchy.
\end{remark}

\section{Regular Languages and Automatic Functions} \label{sec:autofunc}

\begin{remark}[Automatic Functions and Regular Sets]
There is a close relation between what can be computed by register machines
with few registers and regular sets and automatic functions.
Here a function from words to words is called {\em automatic} iff
there is a dfa (deterministic finite automaton)
which can recognise whether the input $x$
and suggested output $y$ match; here the dfa reads both $x$ and $y$ at
the same speed, symbol by symbol. Furthermore, it is assumed that the
input and output are padded with leading zeroes to get that both numbers
have the same length, that is, they are aligned at the back to make the
corresponding digits match (like in the school-book algorithm for adding
decimal numbers). The more frequent way is to align at the
front, but then one has to write all numbers backward in order to avoid
problems; in the present work, it is preferred to write numbers the usual
way and to align at the back.

For a number $x$, let $digits(x,i)$ be the $i$-ary sequence of its digits.
A set $X$ of numbers is regular iff there is a dfa (deterministic
finite automaton) recognising $\{digits(x,i): x \in X\}$ for some $i$
and a function $f$ is automatic iff there are $i,j$ such that the mapping
$digits(x,i) \mapsto digits(f(x),j)$ is automatic as a function from
words to words. Here, it is always assumed that $i \geq 2$ and $j \geq 2$.
The positive result that automatic functions
can be computed by register machines with four registers can even handle
the case that $i \neq j$; however, the choice of $i,j$ requires four registers
if at least one of them is not a power of two, otherwise three are sufficient.
For the result that every function computed by
a machine with one register is automatic, $i=j$ is required,
but any $i \geq 2$ works.

Note that there is a Turing machine model for
computing automatic functions on one-tape Turing machines with one
fixed starting position: Here a function $f$ is automatic iff a
deterministic Turing machine can compute in linear time $f(x)$ from $x$
such that new output $f(x)$ starts at the same position as the old input $x$.
This result also holds when nondeterminsitic machines are used in place of
deterministic ones \cite{CJSS13}.

The first use of the concept of automatic functions and structures dates
back to B\"uchi's work on the decidability of Presburger arithmetic
\cite{Bu60,Bu62}. The notions were formally introduced by
Hodgson \cite{Ho76,Ho83} and, independently, by Khoussainov and
Nerode \cite{KN95}. Gr\"adel \cite{Gr20}
provides the most recent of several surveys in the field \cite{KM10,Ru08}.
The natural numbers with addition, subtraction, comparison and
multiplication by constants forms an automatic structure in which
each of these operations is realised by an
automatic function, see, for example, Jain, Khoussainov, Stephan,
Teng and Zou \cite{JKSTZ14}.

A further example of an automatic function is the function which maps numbers
in ternary to their decimal face value images, here inputs are ternary
and outputs decimal and the corresponding function on the strings is the
identity. So ternary 100 (nine)
would be mapped to one hundred and ternary 121 (sixteen) would be mapped
to one hundred twenty one. Another example is the function which preserves
in decimal numbers the zeroes and maps every other digit $d$ to $10-d$,
for example 102823 to 908287. The latter is automatic as a function from
decimal numbers to decimal numbers, but not as a function from binary
numbers to binary numbers. Also all functions computed by an addition
machines with arbitrarily many registers running for at most constantly
many steps are automatic.
The theorem of Cobham and Semenov implies that functions which are
automatic as functions from binary to binary as well automatic as
functions from ternary to ternary are already definable in the
Presburger arithmetic $({\mathbb Z},+,<,=)$.
\end{remark}

\begin{remark}[Compact Writing of Repetitive Commands]
Multiple identical or similar operations like
\begin{quote}
\begin{itemize}
\item $x=x+y$; $x=x+y$; $x=x+y$;
\item $z=z+v$; $z=z+w$;
\item $u=u+u$; $u=u+u$; $u=u+u$;
\end{itemize}
\end{quote}
will be abbreviated as
\begin{quote}
\begin{itemize}
\item $x=x+3 \cdot y$;
\item $z=z+v+w$;
\item $u=8 \cdot u$;
\end{itemize}
\end{quote}
with the understanding that $3$ and $8$ above are constant and that this is
only done if the result of the first operation goes into the next operation
as above in the case that several operations are in a block. Multiplication
with constants is only possible if this constant is a power of two, as otherwise
the bound on the number of registers on the right side of the assignment
is compromised; Floyd and Knuth had there always two of them and therefore
$u=3 \cdot u$ would need an additional register $t$ with $t=u+u$; $u=u+t$.
However, $u=u+3 \cdot t$; can be realised as $u=u+t$; $u=u+t$; $u=u+t$ and so
while $u=3 \cdot u$ is not permitted in the programs below, $u=u+3 \cdot t$ is
permitted with $u,t$ being different registers. In summary, assignments
of the form
\begin{quote}
\begin{itemize}
\item $x=i \cdot x+j \cdot y+k \cdot z+\ell$;
\end{itemize}
\end{quote}
are allowed provided that $i,j,k,\ell$ are integer constants
and, furthermore, $i$ is either $0$ or a power of $2$ or $-1$
times a power of $2$ and $y,z$ are registers different from $x$
(there might be more or less than $2$ of these registers).
\end{remark}

\begin{remark}[Variables and Constant-ranged operations] \label{rem:varop}
Suppose $k$ is a constant.
The following conventions simplify the writing of programs below.
\begin{quote}
\begin{itemize}
\item The instruction
      $y = k \cdot y$ where $k$ is a constant or a variable (which has constant
      range); here one needs one additional register and just let $z=y$
      and executes $k-1$ times $y=y+z$. If $k$ is a power of two or
      for an instruction of type $x=x+k \cdot y$ no additional register
      is needed.
\item If $0 \leq x < k \cdot y$ then one can by a sequence of instructions load
      the value of $Floor(x/y)$ into the variable $b$ and replaces $x$
      by the remainder of $x/y$. This is done by initially having
      $b=0$ and then $k-1$ copies of the operation:
      \begin{quote}
        if $x \geq y$ then begin let $x=x-y$; let $b=b+1$ end;
      \end{quote} 
      This command is written as $(b,x) = (Floor(x/y),Remainder(x,y))$.
\item Let $\delta$ be the transition function of the deterministic finite
      automaton (dfa) recognising the regular language. In 
      the program in Theorem~\ref{thm:regular}
      below, we can make different copies of the program
      for the constantly many possible values of the 
      bounded variables $a$ and $b$. When there is a need to
      update the values of $a,b$ in 
      instruction 3, the program just jumps to the corresponding
      copy/instruction. Thus, we do not count $a,b$ as needing registers.
\end{itemize}
\end{quote}
These methods will be used in several of the programs.
\end{remark}

\begin{theorem} \label{thm:regular}
A register machine with three registers can check in linear time
whether the $k$-ary representation of a number is a member of
a given regular language, where $k \geq 2$ is constant.
In the case that $k$ is a power of two, only two registers are needed.
\end{theorem}

\begin{proof}
Without loss of generality one assumes that the input $x$ satisfies $x \geq 0$.
The three registers are $x,y,z$ where $x$ holds initially the input
$d_1 d_2 \ldots d_n$ and when entering line 2 of the algorithm below,
the value $d_1 d_2 \ldots d_n 1$ (as a $k$-ary number) which is a coding digit
appended to separate out when the trailing zeroes of the input end and
the new zeroes start which the algorithm appends in subsequent steps.
The register $y$ is initialised as $1$ and after line 2 holds the value
$k^{n+1}$, using the convention that either $d_1 \neq 0$ or $n=0$.
The register $z$ is only used to multiply numbers with the constant $k$
and is needed if $k$ is not a power of $2$.
Furthermore one holds two variables with
constant range, these are $b \in \{0,1,\ldots,k-1\}$ for the current
symbol and $a$ ranges over the possible states of a dfa recognising the language
with constant $start$ being the start state and $accept$ being the set of
accepting states; membership in $accept$ can be looked up in a table
using the value $a$ as an input. The dfa is considered to process the
number from the first digit to the last digit and without loss of generality
one can assume that $\delta(start,0) = start$ and that the dfa never
returns to $start$ after seeing some other digit --- this is to deal with
leading zeroes or the zero itself so that the start is accepting iff $0$
is in the given regular language.
The basic algorithm is the following:
\begin{quote}
\begin{enumerate}
\item Begin read $x$; let $y=1$; let $a=start$; let $x=k \cdot x+1$ using $z$;
\item If $y > x$ then goto 3 \\
      else begin let $y=k \cdot y$ using $z$; goto 2 end; 
\item Let $x=x \cdot k$ using $z$; \\
      if $x\neq y$ then begin let $(b,x) = (Floor(x/y),Remainder(x/y))$; \\
      \mbox{ } \ \ \ let $a = \delta(a,b)$; goto 3 end;
\item If $a \in accept$ then write $1$ else write $0$ End.
\end{enumerate}
\end{quote}
The first line reads an input $x$ being $d_1 d_2 \ldots d_n$ in
the $k$-ary number system, initialises
$y$ as $1$ and appends a digit $1$ at the input obtaining
$d_1 d_2 \ldots d_n 1$. Afterwards in line $2$, $y$ is multiplied with $k$
until it has one digit more than $x$, note that $y=10$ if the input is $0$
and $y=k^{n+1}$ if if the input is positive. An $n$-digit number is at most
$k^n-1$ and therefore $y/k \leq x < y$ after line $2$.

The loop invariants of line 3
after $m$ iterations of the loop, at the start of line 3, are that
$x$ is $d_{m+1} \ldots d_n 1$ times $k^m$ and the digit $d_{m+1}$
is at the position of the power $k^n$ and $a$ is the state
$\delta(start,d_1 d_2 \ldots d_m)$ where $a=start$ in the case that
$m=0$. Here, for every word $w$, $\delta(a,w)$ is the state in which
the dfa is provided that it was first in state $a$ and then read
the digits of the word $w$.

The loop starts with multiplying $x$ by $k$ which makes
the digit $d_{m+1}$ going at the position of $k^{n+1}$.
If $m=n$ then the trailing $1$ goes into the position of $k^{n+1}$
and the loop terminates with $a = \delta(start,d_1 d_2 \ldots d_n)$
which is the correct state of the dfa after reading the $k$-ary
representation of the full number. If the origiinal input is $0$ then
$n=0$, the loop is skipped and $a$ is the start state as required.

If $m<n$ then the loop body determines the value of the variable $b$ as the
smallest number such that, for the current value of $x$,
$x-b \cdot y < y$. At the same time, $y$ is subtracted $b$ times from $x$.
See Remark~\ref{rem:varop} for more details.
Thus after this operation, $x$ has the value 
$d_{m+2} \ldots d_n 1$ times $k^{m+1}$ and $b$ has the value $d_{m+1}$.
After that $a$ is updated: Using the precondition that
$a = \delta(start,d_1 d_2 \ldots d_m)$ before the update,
$a$ is updated to
\begin{quote}
   $\delta(a,d_{m+1}) = \delta(\delta(start,d_1 d_2 \ldots d_m),d_{m+1}) =
   \delta(a,d_1 d_2 \ldots d_m d_{m+1})$
\end{quote}
and so the loop invariant is again true after $m+1$ rounds of the
loop body.

This completes the verification of what is done in Line 3. Line 4
is just the output and the look-up whether $a$ is an accepting
state is trivial.

The program runs in time linear in the number $n$ of $k$-ary digits in $x$,
where the constant factor depends on $k$; the loop in line $2$
increases $y$ by factor $k$ until $y>x$ and the loop body is gone
through $n+1$ times (as $x$ had been multiplied with $k$ in line 1).
The loop in line $3$
moves $k$-ary digit by digit out at the top
until $x=y$, the latter happens as the last $k$-ary digit is $1$
and $y$ is a power of $k$; the loop body is executed $n$ times,
that is, the simulation of the dfa reads $n$ $k$-ary digits and
then checks whether the obtained state is accepting.

Note that the register $z$ was only needed to multiply with $k$ without
overwriting one of the registers $x$ and $y$. If $k=2^\ell$ for some
$\ell \geq 1$, then one does not need this extra register, as one
can replace the multiplication by $k$ with $\ell$ commands
which double up that register.
\end{proof}

\begin{theorem} \label{thm:turing}
One-tape Turing machine constructions running for $f(n)$ steps can be simulated
in $O(f(n)+n)$ steps with three registers in the case that input and output is
binary (or has a base of power of two like octal and hexadecimal numbers)
and with four registers in the case that input and output are $i$-ary
and $j$-ary for arbitrary but fixed $i,j \geq 2$.
\end{theorem}

\begin{proof}
The basic idea is a two-stack simulation of the one-tape Turing machine
which is assumed to be two sided infinite.
The input and output alphabet are assumed to be $\{0,1\}$ and
the tape alphabet is assumed to be of size $2^k$. Furthermore, it is
assumed that the input stands behind the head of the
Turing machine (= is on the right of it)
at the beginning as well as that the output stands before (= is on the
left of) it after the computation. The scrolling takes the Turing
machine only an amount of time linear in input and output size, respectively.
Though the proof is given for $i=2,j=2$, it can be
generalised easily when the input/output alphabet are different,
but one needs one more register to implement the multiplications
when $i$ or $j$ are not powers of $2$.

A three symbol window consisting of
the contents of the cell to the left, under and right of 
the head of the Turing machine is kept in the variables $b, b', b''$.
The parts to the left of
$b$ and to the right of $b''$ are kept in a way similar to
the two-stack simulation of a Turing Machine, but in two
variables (registers) $x$ and $y$ which represent the
portion left of $b$ and right of $b''$ respectively on the tape,
with the cell closest to $b$ (respectively $b''$) being the
highest order $2^k$-ary digit of the tape alphabet.
As the highest order bits might be $0$,
which are not easily recorded in integer variables and in order
read in and out the top position by comparing integers,
an auxillary variable $z$ to indicate top of both stack is used;
it is always a power of $2$.
$z$ being $[2^k]^\ell$ would indicate that there are up to $\ell$-characters
in the variables $x$ and $y$ and there might be trailing blanks
on either side of the input. There is a coding symbol (some constant)
at the bottom of each stack (in order to be able to scroll out the
full content without having to rely on an additional register)
and when this is reached, the register machine knows that there
are only blanks beyond this point on the Turing tape.
Thus whenever in the simulation, one reaches the border
of infinitely many blanks, a new blank is appropriately created in
the variables $x$ or $y$.

The transition function $\delta(a,b,b',b'')$ of the Turing machine 
takes as input current state $a$, the three symbols $b, b',b''$
(symbols on cells left of the head, under the head and right of the head)
and updates them and $c$ to the output $(a,b,b',b'',c)$ which
hold now the new state, the updated values of the three cells
with content denoted as $b,b',b''$
(actually only the cell under the head changes, but this is just for
ease of writing the program) and the move $c$
of the head, which can be left, right, stay or halt.

At the beginning the input to the simulating algorithm is in the
variable $x$ (see line 1 of the algorithm). Then in lines 2, 3 and 4, 
$x$ is copied to $y$ and translated from binary into tape
alphabet and $z$ is appropriately modified (with the resulting value
of $x$ being the empty stack, that is $x=z$), to mimic that
the Turing Machine starts on the input with head being on $b'$ and $b''y$
being the input (with $b''$ being the most significant bit).
Step 5 initializes the start state and sets the variable $x$
as the empty stack representing an infinite sequence of
blanks to the left of $b$.

Step 6 then does a simulation of Turing machine step by step. 
Note that the simulation uses the symbols in base $2^k$.
In each step $z$ is appropriately updated to indicate the top digit
for the variables $x$ and $y$.

After simulation of Turing machine ends, the algorithm goes to
step 7, clears $y$, and then converts the content of $xb$ into
the output $y$ in step 8 --- with $b$ being the most significant bit
and with a translation from the tape alphabet into a binary number.

In the program, the tape alphabet symbol values (constants)
$bitzero$, $bitone$ and $space$ denote codings for $0$, $1$ and blank,
respectively.
The program is as follows.
\begin{quote}
\begin{enumerate}
\item Begin read $x$; let $x=2 \cdot x+1$; let $z=1$;
\item If $z > x$ then goto 3; \\
      let $z=z+z$; goto 2;
\item Let $x=x+x$; let $b=space$; let $b'=space$; let $b''=space$; let $y=z$;
\item If $x = z$ then goto 5;
      let $y = y+b'' \cdot z$; \\
      if $x > z$
      then begin $x=x-z$; $b''=bitone$ end
      else begin $b''=bitzero$ end; \\
      let $x=x \cdot 2^{k+1}$; let $z=z \cdot 2^k$; goto 4; 
\item Let $a=startstate$; let $x=z$;
\item Let $(a,b,b',b'',c)=\delta(a,b,b',b'')$; if $c=halt$ then goto 7; \\
      if $c=left$ then begin $y = y+z \cdot b''$; $b''=b'$; $b'=b$;\\
      \mbox{ } \ if $x \neq z$ then
        begin let $x=2^k \cdot x$;
          let $(b,x) = (Floor(x/z),Remainder(x/z))$ end \\
      \mbox{ } \  else let $b=space$;\\
      \mbox{ } \  let $x=x \cdot 2^k$; let $z=z \cdot 2^k$ end; \\
      if $c=right$ then begin let $x = x+z \cdot b$; let $b=b'$; let $b'=b''$;\\
      \mbox{ } \ if $y \neq z$ then
      begin let $y=2^k \cdot y$;
          let $(b'',y) = (Floor(y/z),Remainder(y/z))$ end \\
      \mbox{ } \ else let $b''=space$;\\
      \mbox{ } \  let $y=y \cdot 2^k$; let $z=z \cdot 2^k$ end; \\
      goto 6;
\item Let $y=0$;
\item If $b=bitone$ then let $y = 2 \cdot y+1$; \\
      if $b=bitzero$ then let $y = 2 \cdot y$; \\
      if $b \notin \{bitzero,bitone\}$ or $x=z$ then goto 9; \\
      let $x=x \cdot 2^k$; if $x=z$ then goto 9; \\
      let $(b,x) = (Floor(x/z),Remainder(x/z))$; goto 8; 
\item Write $y$; End.
\end{enumerate}
\end{quote}
The input/output conventions are such that the program can be written
the most simple way. Other input/output conventions (head on the other
side of input or output) would require scrolling over input/output  
and is here left to the Turing machine program. Time needed 
is $O(n)$ for input processing, $O(f(n))$ for Turing Machine
simulation (and each step of Turing Machine is simulated using
constant number of steps) and $O(n+f(n))$ for the output.
\end{proof}

\noindent
Case, Jain, Seah and Stephan \cite{CJSS13} showed that one can compute
every automatic function with one input length $n$ in $O(n)$ steps on a
one-tape Turing machine where the input and the later produced output
start at the same position. This criterion is even ``if and only if''.
Additional scrolling of the result or the original input as needed for
the Theorem~\ref{thm:turing} is also $O(n)$ steps.
Using this result, one gets the following corollary.

\begin{corollary} \label{cor:turing}
The output of an automatic function with a number interpreted as an $i$-ary
sequence of digits on the input to an output interpreted as an $j$-ary sequence
of digits can be computed by a register machine in linear time using
four registers; if $i$ and $j$ are both powers of two then only three
registers are needed.
\end{corollary}

\begin{corollary} \label{cor:multiautofunc}
Automatic functions with more than one input can be implemented
by a register machine with $5$ registers.
\end{corollary}

\begin{proof}
The idea is to form the convolution of constantly many numbers and
feed this combined input into the automatic function. If two inputs
are given as $i$-ary and $i'$-ary numbers, one forms an $i \cdot i'$-ary
number whose digits are, roughly spoken, pairs of the corresponding
$i$-ary and $i'$-ary digits at the same position.
\begin{quote}
\begin{enumerate}
\item Read first input and translate this input with an automatic function this
      $i$-ary number into
      an $i \cdot i'$-ary number using the same digits, let $x$ denote
      the register holding this number. This translation needs by
      Theorem~\ref{thm:turing} four registers, $x$ being one of them.
      This is like translating the binary number
      $1101$ (thirteen) into the decimal number $1101$ (one thousand
      one hundred and one). Four registers are needed for this.
\item Read the second input and process it using Theorem~\ref{thm:turing}
      with the four registers besides $x$ (which is not modified)
      and let $y$ denote the register which holds the result. This number was
      originally in base $i'$ and is now in base $i \cdot i'$ but
      has the same digits. For example, the quinary number $2323$
      (three hundred thirty eight) becomes the decimal number $2323$
      (two thousand three hundred twenty three).
\item Add $y$ in total $i$ times to $x$. For example,
      $1101$ plus $2323$ plus $2323$ becomes $5747$. This number is
      now the convolution of both numbers, as the first $5$ represents
      the pair $(1,2)$, the next $7$ represents the pair $(1,3)$,
      the next $4$ represents the pair $(0,2)$ and the last $7$
      again represents the pair $(1,3)$.
\item Map with an automatic function this $i \cdot i'$-ary number
      in $x$ to a $j$-ary number. Note that every automatic function
      with constantly many inputs can be viewed as an automatic function
      from the corresponding convolution as a single input to the output.
\end{enumerate}
\end{quote}
If there are three inputs using $i$-adic, $i'$-adic and $i''$-adic
numbers, the automatic functions in the steps before evaluating the
main function translate the $i$-adic, $i'$-adic and $i''$-adic
digits each into $i \cdot i' \cdot i''$-adic digits.
Steps 2 and 3 are repeated for reading and processing the third input,
one adds the corresponding $y$ then $i \cdot i'$ times to $x$.
The so obtained convolution is then mapped to the $j$-adic number
according to the given automatic function to be implemented as
done in the last step of above algorithm.
Analogously one handles even larger amount of inputs. Note that
for automatic functions, the number of inputs is constant, so
there are no loops which create convolutions with an unforeseeable
large base.

As the algorithm consists mainly of executing a constant amount of
automatic functions, it has linear time complexity (where the parameter
$n$ is the longest number of digits of an input) and the constant
factor in the linear term depends not only on the number of inputs
but also on the values of $i,i',i'',\ldots,j,k$ in the simulation of the
automatic functions.
Due to the storage of the so far constructed part $x$ of the convolution,
the simulation needs one register, namely $x$,  more than the worst case
for the simulation of the corresponding automatic function with one input.
If all bases involved are powers of two, the whole algorithm needs only
four registers, see Theorem~\ref{thm:turing} for more details.
\end{proof}

\begin{theorem} \label{thm:thirteen}
If a register program has only one register, one read-statement at
the beginning and one write-statement at the end then it computes
an automatic function $f$ which is of the following special form:

There are integer constants $k,h,i,j,i',j'$ such that $i,i'$ are powers
of $2$ and $k,h > 0$ satisfying the following:
\begin{quote}
\begin{itemize}
\item Either $f(x)=f(x+h)$ for all $x \geq k$ \\
      or $f(x) = ix+j$ for all $x \geq k$ \\
      or $f(x) = -ix+j$ for all $x \geq k$;
\item Either $f(x)=f(x-h)$ for all $x \leq -k$ \\
      or $f(x) = i'x+j'$ for all $x \leq -k$ \\
      or $f(x) = -i'x+j'$ for all $x \leq -k$.
\end{itemize}
\end{quote}
In the second and third line of each item, the computation time is
constant and in the first line it is either constant or exponential or
nonterminating in the number of binary digits $n$ to represent $x$;
furthermore, $f(x)=f(x+h)$ means that either $f$ is undefined on both
inputs (the nonterminating case) or $f(x),f(x+h)$ are both defined 
and equal.
\end{theorem}

\begin{proof}
Note that if there are two different periods $h',h''$ for $x \geq k'$
and $x \leq -k''$ then one can take $k= \max\{k',k''\}$ and $h = h' \cdot h''$.
Thus the theorem can be formulated with just one $k$ and one $h$.

Given a program which uses only one register, one transforms this
program into a normal form with the following steps:

      One introduces a variable $a$ which takes over the sign of $x$
      and has, after this is done, that $x$ is at least $0$
      throughout the program (it is easy to see how to adjust it);
      furthermore one replaces statements of the form ``let $x=x-x$''
      by $x=0$. A statement of the form ``let $x=x+m$;'' where $m$
      is a constant is replaced by ``let $x=x+a \cdot m$'', the reason is
      that the actual value of $x$ is $a \cdot x$ and if $a$ is $-1$,
      then one has to adjust all the comparisons with constants
      and the addition of constants accordingly.

      Furthermore, after adjusting the constants,
      if a statement is ``let $x=m-x$'' then one replaces
      this by the statements
      \begin{quote}
        If $x=m$ then begin let $x=0$; goto line $\ell''$; \\
        If $x=m-1$ then begin let $x=1$; goto line $\ell''$ \\
        If $x=m-2$ then begin let $x=2$; goto line $\ell''$; \\
        $\ldots$\\
        If $x=1$ then begin let $x=m-1$; goto line $\ell''$; \\
        If $x=0$ then begin let $x=m$; goto line $\ell''$; \\
        Let $a=-a$; let $x=x-m$; goto line $\ell''$
      \end{quote}
      Here $\ell''$ is the first line number after this original statement.
      Furthermore, at the end, one replaces the statement ``Write $x$;'' by
      \begin{quote}
         If $a=-1$ then begin let $x=-x$ end; write $x$;
      \end{quote}
      Now in the whole program, the only places where $-x$ occurs instead
      of $x$ are in the beginning before the initial branching and
      at the end where the output is written.

      After that, one expands the program such that every statement
      has its own line and all bounded variables are replaced by having multiple
      copies of the original program and jumping between these when the
      value of the variable changes. If-statements are now only of the
      form ``If $x=m$ then goto linenumber''. Note that all comparisons
      must be with constants as the outcome of comparisons of $x$ with
      itself can be replaced by either a nonconditional jump or by
      completely omitting the statement.

      Now one determines the largest absolute value of a constant $k'$
      which occurs in any statement of the form ``$x=x+i$'' or in
      a comparison; for each $m \leq k'$ and each line number $\ell$,
      one tables out what the output $m'$ would be if $x$ at this line number
      has the value $m$ and then one prefixes the statement in the line
      by a sequence of comparisons with the constants $m$ from $0$ to $k'$
      and then implements the statement
      \begin{quote}
         if $x=m$ then begin let $x=m'$; goto $\ell$ end;
      \end{quote}
      and one puts into the last line $\ell$ only the command ``Write $x$ end.''
      which ends the program. Furthermore, there is a further line
      $\ell'$ for writing the statement ``output is undefined'' in the case 
      that the program does not terminate and then the program stops
      accordingly. One removes after these initial checks all cases handling
      $x \leq k'$ inside this line and has thus a single branch instead
      of an comparison with a constant in the line.
      After that one renumbers the program and flats
      out the if-statements to branch only by jumping to line numbers.

      There are now up to two loops which are taken after the initial
      statements
      \begin{quote}
         if $x \geq 0$ then goto $\ell''$; \\
         let $x=-x$; \\
         goto $\ell'''$;
      \end{quote}
      where each of the loop, if any, is taken in the case of positive and
      of negative numbers. For all $x$ which run through the loop once
      without going below $k'$ at any time, by running through the
      loop, the value of $x$ is 
      either updated to $x+h'$ or $m \cdot x+h'$ where $m$ is a proper power
      of $2$, so $m \in \{2,4,8,16,\ldots\}$; this case means that
      some statements of the form ``let $x=x+x$'' occur in the loop.
      If in the first case $h' \geq 0$
      or in the second case $x \geq h'+k'$ then the loop runs forever
      and the value is undefined; if in the first case $h'<0$ then
      it depends only on the value $Remainder(x/-h')$ which branch
      is taken when $x$ leaves the loop in some line and takes a
      value and then branches to the end. In the case that there is
      no loop, that is, no line in the program is executed twice,
      there are only constantly many statements carried out and for
      large $x$, they are all the same, as never a comparison with
      some constant below $k'$ applies affirmatively. Thus the value
      will be computed by some doubling statements and by statements
      adding or subtracting a constant. Furthermore, the output might
      be multiplied with $-1$ at the end. Thus the value will be of
      the form $i \cdot x+j$ for all sufficiently large positive $x$
      and of the form $i' \cdot x+j'$ for all sufficiently small negative $x$.
      Here $i,i'$ are powers of $2$, perhaps multiplied with $-1$,
      and the $j,j'$ are any integer constants.

      Furthermore, if there
      are two periods $-h'$, $-h''$ then $h$ is the product of these.

      The constant $k$ has to be large enough so that the exceptions which are
      caused by going below $k'$ somewhere in the program or by the fact
      that only, by absolute value, sufficiently large $x$ principally
      leading to infinite loops are also actually leading to infinite loops.
      These considerations together show then that the function is of
      the corresponding form given in the theorem statement.

Automatic functions are closed under composition and finite case distinctions
along regular conditions. Assuming that $x$ is represented in binary or decimal
or by another base (or even by the Fibonacci base which Floyd and Knuth mention)
allows to evaluate the remainder by a fixed number with an automatic function
and the finitely many cases which arise can be piped into a case distinction.
Furthermore, addition and multiplication with constants are automatic functions
in these models, thus the function of the specific form as listed in
theorem statement are indeed automatic.
\end{proof}

\begin{remark} \label{rem:fourteen}
The preceding results show that every function computed
by a register machine with one register is automatic
as a function with respect to any base $d \geq 2$ to represent
the digits. Furthermore, it follows from the work of Cobham \cite{Co69} and
Semenov \cite{Se77} that such partial functions have a graph which is
definable in Presburger arithmetics, that is, which is semilinear.
However, not every semilinear function is computed by a one-register
machine, for example the function $f$ which maps even numbers to $0$ and
odd numbers to itself is definable in Presburger arithmetic but cannot
be computed by a one-register machine, as that would have to check whether $x$
is even and that can only be done by downcounting $x$ in steps of $2$
until a certain constant is reached; however, then the machine has lost all
memory about what $x$ was and cannot output $x$ in the case that $x$ is odd.
A formula defining this function $f$ in Presburger arithmetics is as follows:
$$
   f(x) = y \Leftrightarrow \forall z\,[(x=z+z \rightarrow y=0) \wedge
          (x=z+z+1 \rightarrow y=x)].
$$
On the other hand, all automatic functions with one input
can be computed with four registers in linear time. The fourth register is
only needed in the case that the numbers involved are not represented
in binary, for example if one wants to determine the highest power of $3$
appearing in the ternary representation of the number,
so if $x = 21120$ then the output is $10000$ (both in ternary).
Automatic functions with several inputs can be calculated with five
registers (or even four registers if all inputs and the output of the
automatic function are binary or of other arity which is a power of two).

If an automatic function is based on binary representation and its
range is constant, then one can compute it with a finite automaton where
the state in which it is after processing the full word determines the output;
this automaton runs from the high-order to the low-order bits and needs
only two registers for being implemented in a register machine.

Floyd and Knuth \cite[Theorem 4]{FK90} showed that a register machine with
two registers needs exponentially many steps to compute the greatest common
divisor of $x$ and a constant, say $1$ or $2$, via a program which solves
the greatest common divisor in all cases. If one only wants $x \mapsto
gcd(x,k)$ for constant prime $k$, then this remainder is
computed by membership in a regular set (with respect to binary representation)
and can be carried out with two registers in linear time
(in the addition machine model of this paper which allows the usage of
constants).

Furthermore, one can compute with two registers more semilinear functions
than with one register, for example the automatic function $x \mapsto 3x$.
One can also compute some nonautomatic functions of one input $x$ like
$f(x) = \min\{2^k \cdot \max\{x,1\}: k \geq 1, 2^k \geq x\}$ via
the following program:
\begin{quote}
\begin{enumerate}
\item Begin read $x$; let $y=1$; if $x < 1$ then begin $x=1$ end;
\item Let $x=x+x$; let $y=y+y$; let $y=y+y$; if $y<x$ then goto $2$;
\item Write $x$ End.
\end{enumerate}
\end{quote}
One can show that this function is not automatic; if input and output
are represented in the same base system then the simple argument that
the output of an automatic function can have only constantly many
more symbols than the input is sufficient.

The following two-input function will allow a better separation
by being computed with two registers only in both the model of this
paper and the model of Floyd and Knuth \cite{FK90}. For this, the
second input $y$ is read before the input $x$ in order to check whether
$y$ is strictly positive; if $y \leq 0$ or $x \leq y$ the output is $0$.
\begin{quote}
\begin{enumerate}
\item Begin read $y$; let $x=y-y$; if $y \leq x$ then goto $3$; \\
      read $x$; if $x \leq y$ then begin $x=y-y$; goto $3$ end;
\item Let $x=x+x$; let $y=y+y$; let $y=y+y$; if $y<x$ then goto $2$;
\item Write $x$ End.
\end{enumerate}
\end{quote}
If $y \leq 0$ or $x \leq y$ then the output is $0$ else the output
is the $2^k \cdot x$ for the first number $k\geq 1$ with $2^k \cdot y \geq x$.
Now when fixing $y=1$ and considering $x>y$, the corresponding
restricted function $x \mapsto \min\{2^k \cdot x: k \geq 1, 2^k \geq x\}$
equals to the function $f$ from above on these outputs.
One can now see that for $x$ with $2^{k-1} < x \leq 2^k$, the output
is $2^k \cdot x$. Thus whenever $x>1$ and $f(x+2)-f(x+1)=f(x+1)-f(x)$
then $f(x+1)-f(x)$ is a power of $2$ and thus the set of powers of $2$
in the range of $f$ is first-order definable from $f$, thus assuming
that $f$ is automatic and using that $+,<$ are automatic,
one gets that the powers of $2$ are automatic. Thus the base used
for representing the numbers in the output must be a power of $2$, say $2^j$
\cite{Co69}.
Furthermore, $x$ is a power of $2$ iff $x=1$ or the three conditions $x>1$,
$f(x+1)-f(x) > f(x)-f(x-1)$ are satisfied. So the powers of $2$ are also
definable in the input alphabet and therefore a regular set, again
one can conclude that the base of the digits used in the input alphabet
for the automatic function is a power of $2$, say $2^i$.

Now one shows that $f$ cannot be automatic, as it maps one regular
set to a nonregular set. For this, note that when it comes to a set
$A$ of natural numbers, for all $h$, the set $\{$binary digits of
$a: a \in A\}$ is regular iff the set $\{$digits in base $2^h$ of
$a: a \in A\}$ is regular, as one can consider the digits in base $2^h$
as $h$ subsequent digits in binary at the corresponding positions. For
example, hexadecimal 5F38 is binary $101\,1111\,0011\,1000$ when grouping
the digits accordingly. Now one considers the set $A = \{2^k+1: k \geq 1\}$.
The set of binary representations
of $A$ is the set $\{1\} \cdot \{0\}^* \cdot \{1\}$ which is regular
as this regular expression shows. However, $f(A)$ is the set of all
$2^{k+1} \cdot (2^k+1)$ whose binary representations have
one $1$ at the upper end of the number
and the other $1$ just at the middle, all other digits
are $0$, more precisely, they are of the form
$10^{k-1}10^{k+1}$ when putting the highest order bits first.
An easy application of the pumping lemma shows that the set
of binary representations of $f(A)$ is not regular.

Note that the bound of two registers for this program is optimal,
as Theorem~\ref{thm:thirteen} showed that all register programs
with only one register compute an automatic function.
\end{remark}

\section{Sample Programs} \label{sec:sampleprog}

\noindent
Sample programs are available at
\begin{quote}
  \url{https://www.comp.nus.edu.sg/~fstephan/registermachineprog/}
\end{quote}
and the programs are distinguished from those in the paper as follows:
Output handling and syntax has to follow C/C++; subroutines for the
update of the finite automata state and the checking of the acceptance
conditions are extra C++ functions and the macro for Floor/Remainder
at the regular set program is expanded to the code it stands for.
The programs are the following five:
\begin{enumerate}
\item register-division.cpp \\
      This program implements the division for Open Question (2).
\item register-multiplication.cpp \\
      This program implements the multiplication for Open Question (2).
      Note that squaring (what Floyd and Knuth asked for) is a special
      case of multiplication where both inputs are the same.
\item register-poweroftwo.cpp \\
      This program solves Open Question (5).
\item register-regset-octaleor.cpp \\
      The program checks regular set membership. Here $k=8$, that is,
      $k$ is a power of two. Therefore one can determine regular set
      membership with two registers $x,y$ only.
      The regular set implemented is that where the exclusive or over
      all octal digits of the number gives an octal digit with exactly
      two bits $1$ and one bits $0$. The digits in question are $3$, $5$, $6$
      as they are binary 011, 101, 110.
\item register-regset-threedecimaldigits.cpp \\
      This also checks a membership in a regular set, namely whether there
      are exactly three nonzero decimal digits in the decimal representation
      of the number. The finite automaton counts the nonzero decimal digits
      up to four. As the base is $10$, the program needs three registers
      $x,y,z$.
\end{enumerate}
The programs are in \verb|C++| and in a Unix-type operating system,
one can compile them with the command \verb|g++| and then execute the
file \verb|a.out| created. They are tried to be as similar to the
code in the paper as possible. Note that the programs have implemented
the registers as ``long long int'' but for very large inputs, there might
be overflow errors, either explicit or implicit by computation errors.
Here $u,v,w,x,y,z$ in the variables are registers which can normally
hold arbitrarily large numbers; the variables $a,b,c,d$ are also
implemented as ``long long int'' to avoid type conversion, but they
take only one of finitely many values, for example the current digit
and the state in a finite automaton setting for the last two sample
programs. The programs are there to facilitate the checking of the
validity of the solutions of Open Problems (2) and (5) as well as
of the regular set result.

\section{Conclusion} \label{sec:conclu}

\noindent
Floyd and Knuth investigated the questions which basic numerical operations
can be done by addition machines in linear time and, furthermore, how many
registers are needed for this. The present work answers two problems left open
by Floyd and Knuth completely: In Open Problem (2) they asked whether the
number of five registers to do squaring, integer multiplication and integer
division in linear time can be obtained and the answer is affirmative.
In Open Problem (5) they asked whether one can improve the runtime of
an algorithm to output all powers of two appearing in the binary representation
of a natural number in subquadratic time and the answer is that it can
even be done in linear time.

The authors of the present work think that the usage of constants when
adding, subtracting or comparing is natural and should be allowed. Therefore
their upper bounds on the number of registers obtained deviate in general
by one from those of Floyd and Knuth \cite{FK90}. In order to be honest,
the following table gives the results on the number of registers
needed for the various tasks performed in both models, in the model
of this paper and in the model of Floyd and Knuth.
Floyd and Knuth write ``an integer addition machine can
make use of the constant $1$ by reading that constant into a separate,
dedicated register.'' So the $1$ is given as an additional input to the
machine. All operations with a constant (addition and subtraction) can be
carried out by repeatedly adding or subtracting this register. One can compare
$x$ with constant $k$ by $k-1$ times subtracting the register $y$ which carries
$1$ to $x$, compare with $y$, doing the conditional goto command and then
adding $y$ again $k-1$ times to $x$. Most tasks below need
this additional register. The lower bound proof of Floyd and Knuth \cite{FK90}
for the remainder does not go through in the model of this paper due to
remainder by a constant being doable with two registers, it is just
equivalent to evaluating a finite automaton whose states have the
output value with respect to binary representation (or any other fixed
base) and thus Theorem~\ref{thm:regular} gives bound $2$. However,
Floyd and Knuth asks for an algorithm which can compute the remainder
for all input-output pairs $(x,y)$ and then they prove that in their
model, three registers are needed by showing that determining the remainder
even for any fixed value $y \geq 1$ takes exponential time. The proof
using this fixed value $y$ does not go through, though it seems to the
authors that the result might perhaps hold with another, more sophisticated
proof which is not the topic of this paper, as in the general case a
machine with two registers must forget one of the inputs in order to
store intermediate values, this is generally only possible with the easy
exponential method by continuously subtracting from $x$ or adding to $x$
the value $y$ until $0 \leq x < y$, as the remainders of $x$ by $y$ and
$x-y$ by $y$ and $x+y$ by $y$ are all the same. Floyd and Knuth \cite{FK90}
showed that no method of this type works in linear time.
\begin{center}
\begin{tabular}{|l|l|l|l|l|}
\hline
Operation & Model of & Model of & Bounds of & Source \\
  & this paper & Floyd and & Floyd and & \\
  & & Knuth \cite{FK90} & Knuth \cite{FK90} & \\
\hline
Remainder & $3$ & $3$ & $3$ & Floyd and \\
  & & & & Knuth \cite{FK90} \\
\hline
Greatest Common Divisor & $3$ & $3$ & $3$ & Floyd and \\
  & & & & Knuth \cite{FK90} \\
\hline
Multiplication   & $4$ & $5$ & $6$ & Theorem~\ref{thm:four} \\
\hline
Squaring         & $4$ & $5$ & $6$ & Theorem~\ref{thm:four} \\
\hline
Integer Division & $4$ & $5$ & $6$ & Theorem~\ref{thm:five} \\
\hline
Powers of Two    & $4$ & $5$ & $-$ & Theorem~\ref{thm:six} \\
\hline
Membership in Regular Set & $2$, $3$ & $3$, $4$ & $-$ &
  Theorem~\ref{thm:regular} \\
\hline
Automatic function  & $3$, $4$ & $4$, $5$ & $-$ & Corollary~\ref{cor:turing} \\
(multi-variable)    & $4$, $5$ & $5$, $6$ & $-$ &
  Corollary~\ref{cor:multiautofunc} \\
\hline
Some nonautomatic function & $2$ & $2$ & $3$ & Remark~\ref{rem:fourteen} \\
\hline
\end{tabular}
\end{center}
When there are two entries for regular sets and automatic functions,
it means that the value depends on whether all bases of numbers involved
can be chosen to be a power of $2$. If this is the case then the first value
applies else the second value applies. ``Model of Floyd and Knuth'' means
the best known upper bound so far given (including this paper)
when the program is written in the model of Floyd and Knuth where the
constant $1$ has to be replaced by a designated register and operations
with constants are not allowed.
``Bounds of Floyd and Knuth'' means the best bound on the number of
registers for computing the function in linear time as
it is given in the paper of Floyd and Knuth. The entry ``$-$'' means
in the case of outputting the powers of two that the current time bound
was not matched and in the case of regular sets that the topic was not
investigated by Floyd and Knuth.

``Some nonautomatic function'' in the table means that there is
a function which is not automatic and which can be computed
with two registers in linear time. This is the nonautomatic function in two
variables provided in Remark~\ref{rem:fourteen}. The remainder result
by Floyd and Knuth is a nonautomatic function
which is also computed by a register machine with three registers
in linear time --- note that the divides-relation and thus the
remainder function are nonautomatic in every automatic representation
of natural numbers including those of binary numbers or $k$-ary numbers in
general~\cite{BG04}.

Besides optimising the number of registers used, the paper
also looked into the runtime bounds. Open Problem (5) actually
asked only for the runtime bound and the $O(n)$ algorithm of the
present paper is optimal, as there might be in the worst case $n$ different
numbers which must be output, see the case where the $n$-bit number
is $2^n-1$. Open problems (3) and (4) ask whether algorithms provided
by Floyd and Knuth can be improved asymptotically, that is, for the
obtained time-bounds $O(\log(y) \cdot \log(z))$ steps for (3) and
$O((m+\log(q_1 \cdot q_2 \ldots q_m))\log m)$ steps for (4), can one
replace $O(\ldots)$ by $o(\ldots)$?

The results show that in
both cases this is impossible, provided one uses the Wikipedia
Definition (a) for the multivariate Oh Calculus. However, the
proofs do not work with the alternative Definition (b). Furthermore,
the lower bound for (4) is obtained not for the original task but
for the simpler task to read in $m$ and then $m$ numbers and then read
an index of which number has to be recalled; this simpler task can
always be done at least as fast as sorting. This task fails for
Definition (a) of the Little Oh Calculus but works for the Definition
(b) of the Little Oh Calculus. Thus these two problems are only partially
solved and would be completely solved only when they are answered for
all common versions of the multivariate Oh Calculus; it is even possible
that the answer to the questions depends on the multivariate Oh Calculus
chosen as it did for the simplified task.

There is a close relationship between register machines and automatic
functions. All functions computed by a register machine with only one
register can also be computed by an automatic function and this function
is automatic independently of the base for the representation of the
numbers as a sequence of digits; thus these functions are a proper
subclass of those definable in Presburger arithmetic.
On the other hand, even if one takes
the most general definition for automatic functions from tuples of numbers
to numbers in the sense that for each input and also for the output
one can choose the base of the digits independently of what is taken
for the others of the inputs and the output, the resulting function
can still be computed with five registers in linear time.
Furthermore, there are functions which can be computed with two
registers in linear time which are not automatic. As all the primitive
operations of the register machine (adding, subtracting, comparing)
are automatic functions and relations, one could use the automatic functions
and relations as a primitive operations for a register machine and
would obtain a notion of primitive steps which has the same class
of polynomial time computable functions and relations as the register
machine or as the Turing machine \cite{GJLSS22,Se12}. However,
the more fine-grained time complexities change. Division by constants
and remainders by constant can be done in constant time and
the $\Theta(n)$ bound of Stockmeyer \cite{St76} for register machines
with both addition and multiplication does not carry over to this model.

\medskip
\noindent
{\bf Acknowledgments.}
The authors would like to thank Guohua Wu for detailed explanations
of Cobham's Theorem.

\end{document}